\documentclass[12pt, a4paper,dvipdfmx]{article}
\usepackage{graphicx}
\usepackage{amssymb}
\usepackage{amsmath}
\usepackage{bm}
\usepackage{color}
\usepackage{cite}
\usepackage{epstopdf}            
\usepackage{epsfig}

\newcommand{\beq}{\begin{equation}}
\newcommand{\eeq}{\end{equation}}
\newcommand{\ov}{\overline}

\usepackage[colorlinks=true, linkcolor=black, citecolor=black,
urlcolor=black]{hyperref}

\begin{document}

\begin{titlepage}

\begin{flushright}

IPMU16-0095

KIAS-PREPRINT-P16050

\end{flushright}

\vskip 2cm
\begin{center}

{\Large
{\bf 
Flavor physics induced by light $Z'$ from SO(10) GUT
}
}

\vskip 2cm

Junji Hisano$^{1,2,3}$,
Yu Muramatsu$^{4,5}$,
Yuji Omura$^{1}$
and
Yoshihiro Shigekami$^{2}$

\vskip 0.5cm

{\it $^1$
Kobayashi-Maskawa Institute for the Origin of Particles and the
Universe, \\ Nagoya University, Nagoya 464-8602, Japan}\\[3pt]
{\it $^2$Department of Physics,
Nagoya University, Nagoya 464-8602, Japan}\\[3pt]
{\it $^{3}$
Kavli IPMU (WPI), UTIAS, The University of Tokyo, Kashiwa, \\ Chiba 277-8583, Japan}\\[3pt]
{\it $^{4}$ School of Physics, KIAS, Seoul 130-722, Republic of Korea}\\[3pt]
{\it $^{5}$ Quantum Universe Center, KIAS, Seoul 130-722, Republic of Korea}

\vskip 1.5cm

\begin{abstract}

In this paper, we investigate predictions  of the SO(10) Grand Unified Theory (GUT), where an extra U(1)$^\prime$ gauge
symmetry remains up to the supersymmetry (SUSY) breaking scale. The minimal setup of SO(10) GUT unifies quarks and leptons into a ${\bf 16}$-representational field in each generations. 
The setup, however, suffers from the realization of the realistic Yukawa couplings at the electroweak scale.
In order to solve this problem, we introduce ${\bf 10}$-representational matter fields, and then the two kinds of matter fields mix with each other at the SUSY breaking scale, where the extra U(1)$^\prime$ gauge symmetry breaks down radiatively.
One crucial prediction is that the Standard Model quarks and leptons are given by the linear combinations
of the fields with two different U(1)$^\prime$ charges. The mixing also depends on the flavor.
Consequently, the U(1)$^\prime$ interaction becomes flavor violating, and the flavor physics is the smoking-gun signal of our GUT model. The flavor violating $Z'$ couplings are related to the fermion masses and the CKM matrix, so that
we can derive some explicit predictions in flavor physics. We especially discuss $K$-$\overline{K}$ mixing, 
$B_{(s)}$-$\overline{B_{(s)}}$ mixing, and the (semi)leptonic decays of $K$ and $B$ in our model.
We also study the flavor violating $\mu$ and $\tau$ decays and discuss the correlations among the physical observables in this SO(10) GUT framework.
\end{abstract}

\end{center}
\end{titlepage}

%%%%%%%%%%%%%%%%%%%%%%%%%%%%%%%%%%
\section{Introduction}
\label{sec:intro}
%%%%%%%%%%%%%%%%%%%%%%%%%%%%%%%%%%
The supersymmetric SO(10) Grand Unified Theory (GUT) is one of the promising candidates for
the underlying theory of the Standard Model (SM). The GUT elegantly explains the origin of the 
SM gauge groups and shows that the SM matter fields can be unified into three-family ${\bf 16}$-representational
fields in the minimal SO(10) GUT \cite{so10}. 
In fact, several problems have been pointed out in the framework of the minimal setup, but
the supersymmetric GUT deserves to be believed because of the beauty and the elegant explanations of the origins of
not only the SM gauge groups but also the electroweak (EW) scale, so that 
a lot of solutions for the problems have been also proposed so far.

For instance, the unification of the SM matters, i.e. the unification of the Yukawa couplings, is a very attractive hypothesis, but unfortunately the precise experimental measurements of the masses and the CKM matrix require some deviation from the unified Yukawa couplings. One simple solution is to add higher-dimensional operators involving Higgs fields to break the SO(10) and SU(5) gauge symmetries  \cite{FermionMass-via-HDO}.\footnote{Introducing additional Higgs fields \cite{FermionMass-via-EH} and additional matter fields \cite{Barr:1981wv} have been proposed so far.} 
In the minimal SO(10) GUT, there is only up-type Yukawa coupling, $h_{ij}$, at the renormalizable level, but realistic Yukawa couplings could be effectively obtained by including such a higher-dimensional operator contribution. However, we have to assume that the additional contributions and $h_{ij}$ are compatible and cancel each other, in order to realize the large mass hierarchy between top and bottom quarks, if $\tan \beta$ is small. $h_{tt}$, which corresponds to the top quark mass is ${\cal O}(1)$, and then the effective term should be also ${\cal O}(1)$ for the bottom quark mass.

Another issue is how to achieve the Higgs mass observed around 125 GeV. In the supersymmetric GUT,
the EW scale is naturally derived and the lightest Higgs mass is predicted.
The lower bound on the predicted Higgs mass is roughly the $Z$ boson mass and shifted by
the supersymmetry (SUSY) breaking scale. In order to achieve the 125 GeV mass, 
it is known that the SUSY breaking scale should be ${\cal O}(100)$ TeV \cite{HSSUSY}, unless the SUSY spectrum is unique (e.g. see Refs. \cite{Omura,Abe:2012xm,Feng:2012jfa,Baer:2012up}). 
In this high-scale SUSY scenario, however, 
the problem about the Yukawa couplings is revived because such ${\cal O}(100)$ TeV SUSY scale requires small $\tan \beta$ for the 125 GeV Higgs mass. Thus, we have to consider some mechanisms to 
realize the large mass hierarchy especially between top and bottom quarks, in order to avoid 
the remarkably large coefficients of higher-dimensional operators.

In Ref. \cite{Hisano-so10}, the authors propose an extension of the minimal SO(10) GUT to explain the hierarchy
in the high-scale SUSY scenario. In addition to the ${\bf 16}$ matter fields, three-family ${\bf 10}$ fields
are introduced and the realistic Yukawa couplings are achieved by the mixing between two kinds of SU(5) ${\bf \bar 5}$-representational fields originated from ${\bf 16}$ and ${\bf 10}$ fields respectively.
An interesting point is that $Z'$ interaction, predicted by SO(10) gauge symmetry, becomes flavor-dependent
because the SU(5) ${\bf \bar 5}$-representational fields carry different U(1)$^\prime$ charges \cite{Hisano-so10}.
Once we assume that U(1)$^\prime$ is radiatively broken at the SUSY scale as the EW scale is,
the flavor violating processes triggered by $Z'$ are verifiable in the flavor experiments, such as the LHCb, the Belle II, the COMET and the Mu2e experiments.\footnote{Introduction of additional matter multiplets at low energy enhances proton decay by $X$-boson exchange, since the gauge coupling constants at the GUT scale become larger \cite{protondecay}. If proton decay is discovered, embedding quarks and leptons to GUT multiplets may be resolved. 
%
%  \bf
%
}

In this paper, we investigate our predictions of the flavor violating couplings quantitatively
and discuss the flavor violating processes relevant to our SO(10) GUT.
Especially, all elements of our Flavor Changing Neutral Currents (FCNCs) involving Z$^\prime$ 
become large so that we should carefully check the consistencies with the observables related 
to the first and second generations: $K$-$\overline{K}$ mixing and lepton flavor violating $\mu$ decays.
Besides, we find that the $(b, \, s)$ element of the $Z'$ couplings tends to be larger than the others because of the fermion masses,
as we will see in Sec. \ref{sec2-2}. Then, we study $B$ physics as well: $B_{(s)}$-$\overline{B_{(s)}}$ mixing,  $B_{(s)} \to \mu^+ \mu^-$ and so on. We also show our prediction on $K_L \to \pi \nu \overline{\nu}$ motivated by the KOTO experiment.
Then, we discuss lepton flavor violations (LFV) in our model.
Interestingly, we could find some correlations between the observables of mesons and leptons.
Then, we show our predictions for $\mu \to 3 e$ and the $\mu$-e conversion in nuclei.

Our paper is organized as follows.
In Sec. \ref{sec2}, we give a short review on our setup, based on Ref. \cite{Hisano-so10}.
Then, we show how well the realistic Yukawa couplings can be achieved and
discuss our prediction of the Z$^\prime$ FCNCs in Sec. \ref{sec2-1}.
In Sec. \ref{sec3}, we study flavor physics in our SO(10) GUT, concentrating on
the relevant processes: $K$-$\overline{K}$ mixing, $B_{(s)}$-$\overline{B_{(s)}}$ mixing, $\mu \to 3 e$, 
and so on. We give some analyses on $\Delta F=1$ processes as well, but
we will conclude that $\epsilon_K$ gives the strongest bound on our model.
We also show the correlation between $\epsilon_K$ and LFV $\mu$ decays: 
$\mu \to 3 e$ and $\mu$-$e$ conversion in nuclei in Sec. \ref{sec3-2}.
Then, we see that our model could be tested at the COMET and the Mu2e experiments near future.
Finally, we present some results for LFV $\tau$ decays in Sec. \ref{sec3-4}.
Sec. \ref{sec5} is devoted to summary.

\section{Overview of the setup}
\label{sec2}
In the minimal setup of the SO(10) GUT, the matter superfields belong to
the ${\bf 16}$ representation and the Yukawa couplings are described by one $3 \times 3$ matrix, $h_{ij}$:
\beq
W_{\rm{min}}=h_{ij} {\bf 16 }_i  {\bf 16}_j {\bf 10}_H,
\eeq
where $i, \,j=1, \, 2, \, 3$ denote the generations and ${\bf 10}_H$ is the chiral superfield for the Higgs.
$ {\bf 16 }_i$ includes all quarks and leptons in each generations, so that it is hard
for this minimal setup to describe the mass hierarchies in the each sectors and the CKM matrix.
 
In Ref. \cite{Hisano-so10}, the authors propose a simple setup of the SO(10) GUT to
realize the realistic Yukawa couplings at the EW scale. We introduce three ${\bf 10}$-representational 
chiral superfields (${\bf 10}_i$) in addition to ${\bf 16 }_i$. Then we write down the additional Yukawa couplings 
and mass terms for ${\bf 10}_i$:
\beq
W_{\rm{ex}}=g_{ij} {\bf 16}_i {\bf 10}_j {\bf 16}_H+ 
\mu_{10\,ij} {\bf 10}_i  {\bf 10}_j.
\eeq
${\bf 16}_H$ is an extra Higgs field to break the remaining U(1)$^\prime$ symmetry.
In order to sketch our idea, let us focus on the down-type quark sector,
assuming that SO(10)-adjoint chiral superfields, ${\bf 45}_H$ and ${\bf 45}'_H$, break SO(10)
to G$_{\rm SM}$ $\times$ U(1)$^\prime$ at the GUT scale.
There are two kinds of right-handed down-type quarks which carry different U(1)$^\prime$ charges, after the symmetry breaking: $d^{(16)}_{L,R \, i}$, and $d^{(10)}_{L,R \, i}$, which are originated from the ${\bf 16}_i$
and ${\bf 10}_i$. Involving the scalar component ($\Phi$) of the SM singlet in ${\bf 16}_H$,
we find the $6 \times 6$ mass matrixes for the down-type quarks induced by $W_{\rm{min}}+W_{\rm{ex}}$:
\beq
{\cal L}_d = -  \overline{ \begin{pmatrix}d^{(16)}_{R \, i} & d^{(10)}_{R \, i} \end{pmatrix} } \begin{pmatrix} h_{ij} v_d  &  g_{ij} \langle \Phi \rangle  \\
0 &  \mu_{10\,ij}  \end{pmatrix}   \begin{pmatrix} d^{(16)}_{L \, j} \\  d^{(10)}_{L \, j} \end{pmatrix},
\label{eq;massmatrix-down}
\eeq
where $v_d$ denotes the nonzero VEV of the down-type Higgs doublet belonging to ${\bf 10}_H$.
As we see in Eq. (\ref{eq;massmatrix-down}), if $\Phi$ develops nonzero VEV, $d^{(16)}_i$ and $d^{(10)}_i$ mix with each other and the lightest three down-type quarks can be interpreted as the SM down-type quarks.
Note that $\Phi$ is charged under U(1)$^\prime$, so that non-vanishing VEV of $\Phi$ spontaneously breaks U(1)$^\prime$.

Let us define the mixing as follows: 
\beq
\label{eq;mixing}
\begin{pmatrix} d_R \\ d^h_R \end{pmatrix}=  U_d  \begin{pmatrix} d^{(16)}_R \\ d^{(10)}_R \end{pmatrix}= \begin{pmatrix} \Hat U^d_{16} & \Delta U_d \\ \Delta U'_d & \Hat U^d_{10} \end{pmatrix} \begin{pmatrix} d^{(16)}_R \\ d^{(10)}_R \end{pmatrix}, 
\eeq
where $d_R$ is the right-handed SM quark and $d^h_R$ is the extra heavy quark.
In Eq. (\ref{eq;mixing}), the flavor index, $i$, is omitted. $U_d $ is a $6\times 6$ unitary matrix,
and $\Hat U^d_{16,10}$ and $\Delta U_d^{(\prime)}$ are $3 \times 3$ matrices
that satisfy, for instance, 
\beq
\label{eq;unitary}
(\Hat U^{d }_{16} )_{ik} (\Hat U^{d *}_{16})_{jk} + (\Delta U_d)_{ik} (\Delta U^*_{d })_{jk} =\delta_{ij}.
\eeq
The mixing unitary matrix, $U_d$, is fixed by the parameters in the $W_{\rm{ex}}$, 
following Eqs. (\ref{eq;massmatrix-down}) and (\ref{eq;mixing}).
Now, let us simply consider the mixing in the limit that $h_{ij} v_d$ are much smaller 
than $g_{ij} \langle \Phi \rangle $ and $ \mu_{10\,ij}$. Then, the left-handed SM quarks are given by $d^{(16)}_{L \, i} (\equiv d_{L \, i})$. The mixing for the right-handed quarks is given by the equation,
\beq
 (\Hat U^{d }_{16})_{ik} g_{kj} \langle \Phi \rangle + (\Delta U_d)_{ik}  \mu_{10\,kj}=0.
\eeq

Using the $\Hat U^d_{16}$ parameters, the Yukawa couplings ($h^d_{ij}$) to generate the SM down-type quark mass matrix is given by 
\beq
\label{eq;down-quarkYukawa0}
h^d_{ij} = (\Hat U^d_{16})_{ik} h_{kj}. 
\eeq
$h_{ij}$ is expected to explain the up-type SM quark mass matrix, so that $\Hat U^{d }_{16}$ matrix
should be fitted to realized the mass hierarchy between the up-type and down-type quarks.
However, it is difficult for $h^d_{ij}$ to be realistic because of the relation in Eq. (\ref{eq;unitary}).
The elements of $\Hat U^{d }_{16}$ could be ${\cal O} (1)$, but cannot be too large because of the unitary condition.
As discussed in Ref. \cite{Hisano-so10}, the mass hierarchy between top and bottom quarks can be achieved,
but the other mass relations and  the CKM matrix  especially involving the first and second generations require
too large $(\Hat U^{d }_{16})_{ij}$, because of the very light up quark mass.
In order to complement the suppression factors, one can introduce higher-dimension operators involving 
${\bf 45}_H$ and ${\bf 45}'_H$ fields  and modify the relation in Eq. (\ref{eq;down-quarkYukawa0}) as
\beq
\label{eq;down-quarkYukawa1}
h^d_{ij} = (\Hat U^d_{16})_{ik} (h^u_{kj} + \epsilon \,c^d_{kj}). 
\eeq
$ \epsilon$ denotes the suppression factor from the ratio between the VEVs of ${\bf 45}_H$ and ${\bf 45}'_H$ and the unknown cut-off scale where the higher-dimensional operators are induced.
$h^u_{ij}$ are the Yukawa couplings for the up-type SM quarks and slightly deviated from $h_{ij}$, because of the higher-dimensional operators. 
$c^d_{ij}$ are the free parameters in our model, and assumed to be ${\cal O}(1)$.

In the same manner, we can discuss the lepton sector.
If the SU(5) relation is respected approximately, the Yukawa couplings ($h^l_{ij}$) for the charged lepton masses
are given by $h^d_{ij}$. The experimental results, however, require slight SU(5) symmetry breaking effects. Then we introduce 
\beq
\label{eq;leptonYukawa1}
h^l_{ij} = (\Hat U^l_{16})_{ik} (h^u_{kj} + \epsilon \,c^l_{kj}),
\eeq
where $\Hat U^l_{16}$ is the $3 \times 3$ matrix which satisfies the relation in Eq. (\ref{eq;unitary}),
replacing $d$ with $l$. In principle, $\Hat U^d_{16}$ and $\Hat U^l_{16}$ ($c^d_{ij}$ and $c^l_{ij}$) are different from each other, because the effective couplings generated by the VEVs of ${\bf 45}_H$ and ${\bf 45}'_H$ are different. We could expect that the corrections of the higher-dimensional operators 
are sufficiently small in the effective $g_{ij}$ and $\mu_{10 \, ij}$ couplings, and then 
it would be reasonable to assume 
\beq
\label{eq;SU(5)relation}
(\Hat U^l_{16})_{ij} \simeq (\Hat U^d_{16})_{ij}.
\eeq
In this case, the realistic Yukawa couplings are achieved by $\epsilon \, c^{d,l}_{ij}$.

\subsection{Requirements for the realistic Yukawa couplings}
\label{sec2-1}
The up-type quark Yukawa couplings $h^u_{ij}$ are defined as follows, without loss of generality:
\beq
h^u_{ij}= \frac{m^u_{i}}{v^u} \delta_{ij}, \label{eq;mixingandhighdimu}
\eeq
where $v^u$ is the VEV of the up-type Higgs doublet and $m^u_i$ are the up-type quark masses, respectively.
According to Eqs. (\ref{eq;down-quarkYukawa1}) and (\ref{eq;leptonYukawa1}), we find the equations which should
be satisfied by the mixing parameters and coefficients of higher-dimensional operators:
\begin{eqnarray}
h^d_{ij}&=&\frac{m^d_{i}}{v^d} \, (V^*_{CKM})_{ji} = (\Hat U^d_{16})_{ik} \left (  \frac{m^u_{k}}{v^u}  \, \delta_{kj} + \epsilon \,c^d_{kj} \right ), \label{eq;mixingandhighdimd} \\
h^l_{ij}&=&\frac{m^l_{i}}{v^d} \, (V^*_{R})_{ji} = (\Hat U^l_{16})_{ik} \left(  \frac{m^u_{k}}{v^u}  \, \delta_{kj} + \epsilon \,c^l_{kj} \right ), \label{eq;mixingandhighdiml}
\end{eqnarray} 
where $v^d$ is the VEV of the down-type Higgs doublet and $m^{d}_{i}$ ($m^{l}_{i}$) are the down-type quark (lepton) masses, respectively. $V_R$ is the unitary matrix and identical to the CKM matrix ($V_{CKM}$) in the SU(5) limit. 
The other constraints on the matrices, $\Hat U^{d,l}_{16}$ and $c^{d,l}$, are from Eq. (\ref{eq;unitary}) and the purturbativity.

Note that heavy modes are integrated out around the U(1)$^\prime$ breaking scale, 
and then $h^{d,l}_{ij}$ in Eqs. (\ref{eq;down-quarkYukawa1}) and (\ref{eq;leptonYukawa1})
are generated. In order to compare our predictions with the observed values of quark and lepton masses and CKM matrix,
we need include the RG corrections from the U(1)$^\prime$ breaking scale (${\cal O} (100)$ TeV) to the low scale, e.g. the EW scale ($M_Z$).
 
We evaluate the realistic Yukawa couplings at the U(1)$^\prime$ breaking scale ($M_{Z'}$) from the central values of 
the experimental measurements summarized in Table \ref{table;input}.
There are three scales relevant to our scenario: $M_Z$, gluino mass (around 1 TeV), and $M_{Z'}$. 
First, we evolve the input parameters in Table \ref{table;input} into the ones at the $M_Z$ scale.
We use Mathematica package RunDec \cite{RunDec} to evaluate the running quark masses.
We translate lepton pole masses to $\overline{\text{MS}}$ running masses at the $M_Z$ scale, following Ref. \cite{SMrun}.
In our analysis, the up-type Yukawa coupling is defined as the diagonal form at $M_Z$, using the up-type quark masses.
The down-type Yukawa coupling is given by the CKM matrix and the down-type quark.\footnote{In fact, we can multiply arbitral unitary matrices to define the Yukawa couplings.
When we match our predictions with the realistic Yukawa couplings, we do not take such degrees of freedom into account.}
Next, we drive the Yukawa matrices from the $M_Z$ scale to $1$ TeV, using the SM RG running at the two-loop level \cite{SMrun}. We assume that all gaugino mass reside around 1 TeV, so that we convert the $\overline{\text{MS}}$ scheme
into the $\overline{\text{DR}}$ scheme at 1 TeV according to Ref. \cite{MStoDR} and drive the Yukawa matrices from $1$ TeV scale to $100$ TeV scale, including the gaugino contributions.
In our scenario, the other SUSY particles reside around 100 TeV.
As a result, we obtain the following values at 100 TeV:
\begin{eqnarray}
(m^u_i) &=&(8.4 \times 10^{-4} \, \text{GeV},\, 0.43 \, \text{GeV},\,
1.2 \times 10^2 \, \text{GeV})\, ,\nonumber  \\
(m^d_i) &=&(1.9 \times 10^{-3} \, \text{GeV} ,\,3.8 \times 10^{-2} \, \text{GeV},\,
1.9 \, \text{GeV} )\,,\nonumber  \\
(m^l_i) &=&(5.0 \times 10^{-4} \, \text{GeV},\,0.11 \, \text{GeV} ,\,
1.8  \, \text{GeV} )\,,  \label{eq;mass100TeV}
\end{eqnarray}
and
\begin{equation}
V_{CKM}=\left( \begin{array}{ccc}
9.7 \times 10^{-1} & 
2.3 \times 10^{-1} & 
1.5 \times 10^{-3} - 3.6 \times 10^{-3} i \\
-2.3 \times 10^{-1} - 1.6 \times 10^{-4} i &
9.7 \times 10^{-1} &
4.4 \times 10^{-2} \\
8.5 \times 10^{-3} -3.5 \times 10^{-3} i &
-4.3 \times 10^{-2} - 8.2 \times 10^{-4} i &
1.0 \\
\end{array} \right). \label{eq;CKM100TeV} \end{equation}
Note that the quark and lepton masses, $m^f_i$ ($f=u, \, d, \, l$), at 100 TeV are obtained, multiplying 
 the running Yukawa couplings by $v=174$ GeV. 
$h^f_{ij}$ at 100 TeV are given by Eqs. (\ref{eq;mixingandhighdimu}), (\ref{eq;mixingandhighdimd}), and  (\ref{eq;mixingandhighdiml}), taking $\tan \beta$ into account.
In the next subsection, $(\Hat U^d_{16})_{ij}$ and $(\Hat U^l_{16})_{ij}$
are calculated, using the obtained $h^{f}_{ij}$ and the relations in Eqs. (\ref{eq;mixingandhighdimd}) and  (\ref{eq;mixingandhighdiml}).

%%%%%%%%%%%%%%%%%%%%%%%%%%%%%%%%
%%%%%%%%Input parameters%%%%%%%%%%%%%%%%
\begin{table}
\begin{center}
  \begin{tabular}{|c|c||c|c|} \hline
  $m_e$ & 0.5110 MeV \cite{PDG} &  $\lambda$& 0.22543$^{+0.00042}_{-0.00031}$  \cite{CKMf15}    \\ 
  $m_\mu$ & 105.7 MeV  \cite{PDG}  & $A$& 0.8227$^{+0.0066}_{-0.0136}$  \cite{CKMf15}  \\ 
  $m_\tau$ & 1.777 GeV  \cite{PDG}  &  $\overline{\rho}$& 0.1504$^{+0.0121}_{-0.0062}$  \cite{CKMf15} \\ 
  $m_d$(2 GeV) & 4.8$^{+0.5}_{-0.3}$ MeV \cite{PDG} &  $\overline{\eta}$& 0.3540$^{+0.0069}_{-0.0076}$  \cite{CKMf15} \\ 
    $m_s$(2 GeV) & 95$\pm 5$ MeV \cite{PDG} & $M_Z$ & 91.1876(21) GeV  \cite{PDG}  \\ 
     $m_{b}(m_b)$&4.18$\pm 0.03$ GeV  \cite{PDG}  &  $M_W$ & $80.385(15)$ GeV  \cite{PDG}   \\ 
   $\frac{2m_{s}}{(m_u + m_d)}$(2 GeV)& 27.5$\pm 1.0$ \cite{PDG} & $\sin^2 \theta_W$ & 0.23126(5)  \cite{PDG}        \\ 
      $m_{c}(m_c)$&1.275$\pm 0.025$ GeV  \cite{PDG}  & $G_F$  & 1.1663787(6)$\times 10^{-5}$ GeV$^{-2}$  \cite{PDG}   \\
      $m_t$& 173.21$\pm 0.51 \pm 0.71$ GeV  \cite{PDG}  &  $\alpha$ & 1/137.036  \cite{PDG}  \\ 
         && $\alpha_s(M_Z)$ & $0.1193(16)$ \cite{PDG}        \\ \hline
  \end{tabular}
 \caption{The input parameters in our analysis. The CKM matrix, $V_{CKM}$, is written in terms of $\lambda$, $A$, $\overline{\rho}$ and $\overline{\eta}$ \cite{PDG}.}
  \label{table;input}
  \end{center}
\end{table}
%%%%%%%%%%%%%%%%%%%%%%%%%%%%%%%%
%%%%%%%%%%%%%%%%%%%%%%%%%%%%%%%%

%%%%%%%%%%%%%%%%%%%%%%%%%%%%%%%%%%%%%%%%%%%%%%
%%%%%%%%%%%%%%%%%%%%%%%%%%%%%%%%%%%%%%%%%%%%%%

\subsection{Flavor violating $Z'$ couplings}
\label{sec2-2}
As we see in Eq. (\ref{eq;mixing}), the SM right-handed down-type quarks and left-handed leptons
are given by the linear combinations of the parts of ${\bf 16}_i$ and  ${\bf 10}_i$ in the SO(10) GUT.
We consider the scenario that an extra U(1)$^\prime$ symmetry remains up to the SUSY breaking scale.
Then, we find that the particles from ${\bf 16}_i$ and  ${\bf 10}_i$ carry different U(1)$^\prime$ charges 
corresponding to the representations of SO(10).
In fact, the U(1)$^\prime$ charges of $d^{(16)}_{R \, i}$ and $d^{(10)}_{R \, i}$ are $-3$ and $2$, respectively,
and the ones of $l^{(16)}_{L \, i}$ and $l^{(10)}_{L \, i}$ are $3$ and $-2$ \cite{Hisano-so10}.
The U(1)$^\prime$  symmetry breaking is triggered by the nonzero VEV of $\Phi$, and causes the mixing between the
different-U(1)$^\prime$-charged fields. Consequently, the $Z'$ interaction becomes flavor violating as follows: 
\begin{equation}
\label{eq;gauge}
{\cal L}_g =g' \Hat{Z}'_{\mu} \left (A^l_{ij}\overline{l^i_L} \gamma^\mu l^j_L-A^d_{ij}\overline{d^i_R} \gamma^\mu d^j_R-\overline{q^i_L}\gamma^\mu q^i_L+\overline{u^i_R}\gamma^\mu u^i_R+\overline{e^i_R}\gamma^\mu e^i_R \right ), 
\end{equation}
where $q^i_L$, $u^i_R$ and $e^i_R$ are the mass eigenstates of the left-handed quarks, right-handed up-type quarks and right-handed charged leptons. Note that $\Hat{Z}'_{\mu}$ is not the mass eigenstate. This mixes with the Z boson, as mentioned below. $A^{l,d}_{ij}$ are given by
\beq
A^{d}_{ij}=5 (\Hat U^{d }_{16} )_{ik} (\Hat U^{d }_{16} )^*_{jk}-2 \delta_{ij},~A^{l}_{ij}=5 (\Hat U^{l }_{16} )^*_{ik} (\Hat U^{l }_{16} )_{jk}-2 \delta_{ij}.
\label{eq;FVC}
\eeq
Assuming the SU(5) relation in Eq. (\ref{eq;SU(5)relation}), $A^{d}_{ij}$ and $A^{l}_{ij}$ satisfy 
\beq
\label{eq;SU(5)relation2}
A^{d}_{ij} \simeq (A^{l}_{ij})^*.
\eeq

\begin{figure}[!t]
\begin{center}
{\epsfig{figure=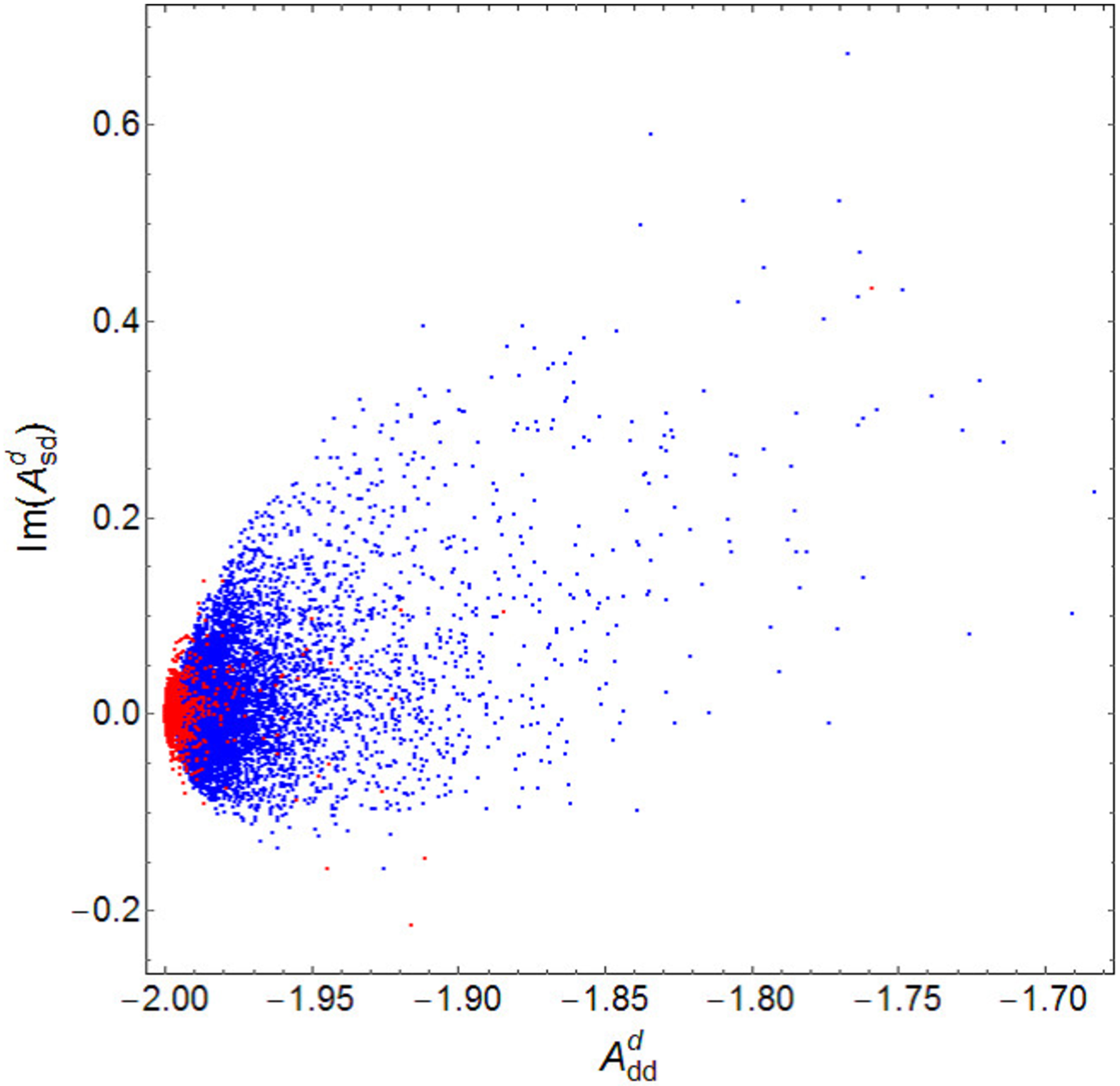,width=0.45\textwidth}}\hspace{0.5cm}{\epsfig{figure=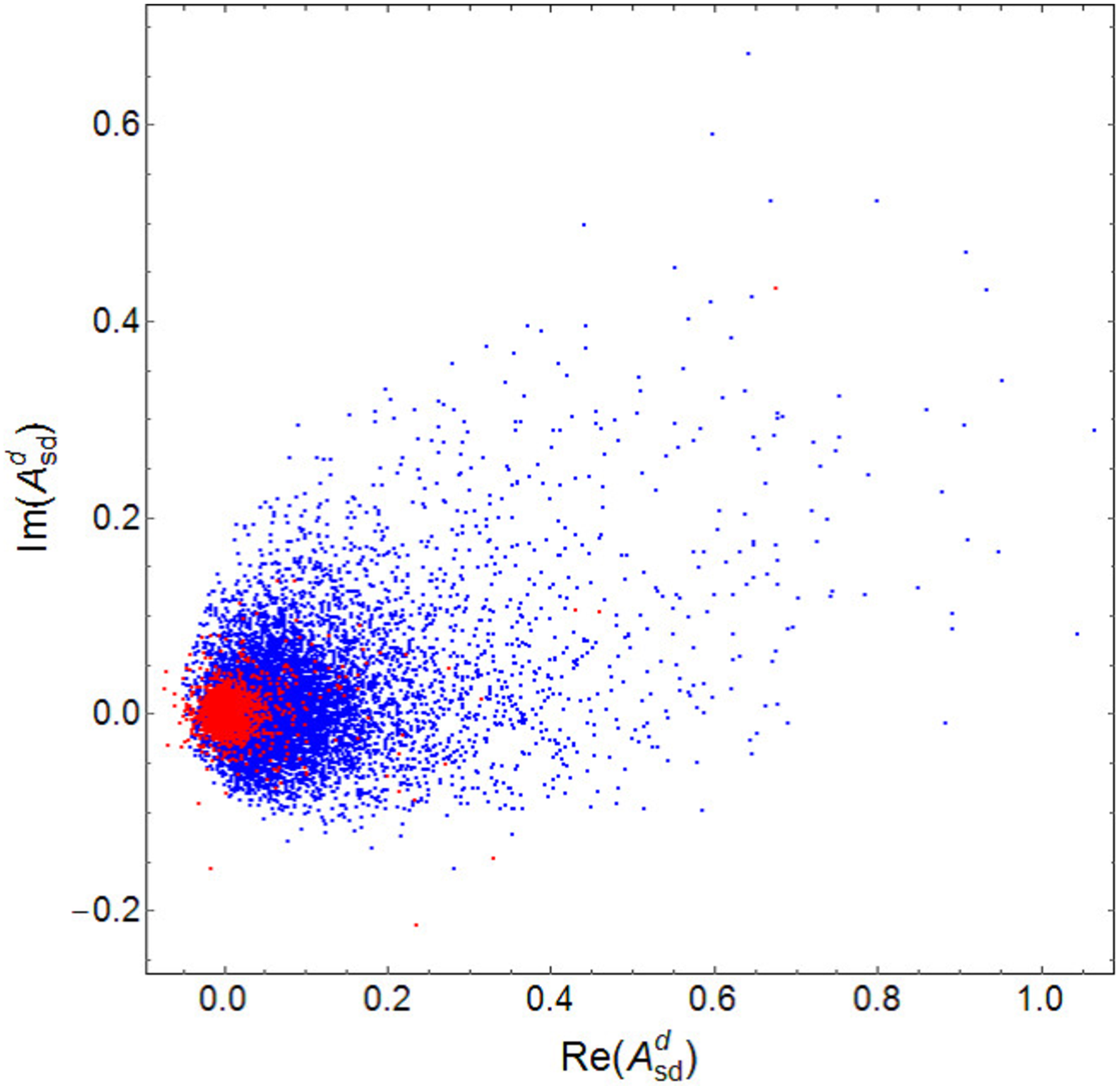,width=0.45\textwidth}}
\caption{ Our predictions for $A^d_{dd}$ (left) and $A^d_{sd}$ (right). The coefficients of higher-dimensional operators satisfy $|\epsilon \,c^d_{ij}|<10^{-2}$ (red) and $|\epsilon \,c^d_{ij}|<10^{-3}$ (blue).}
\label{fig;FCNC1}
\end{center}
\end{figure}

\begin{figure}[!t]
\begin{center}
{\epsfig{figure=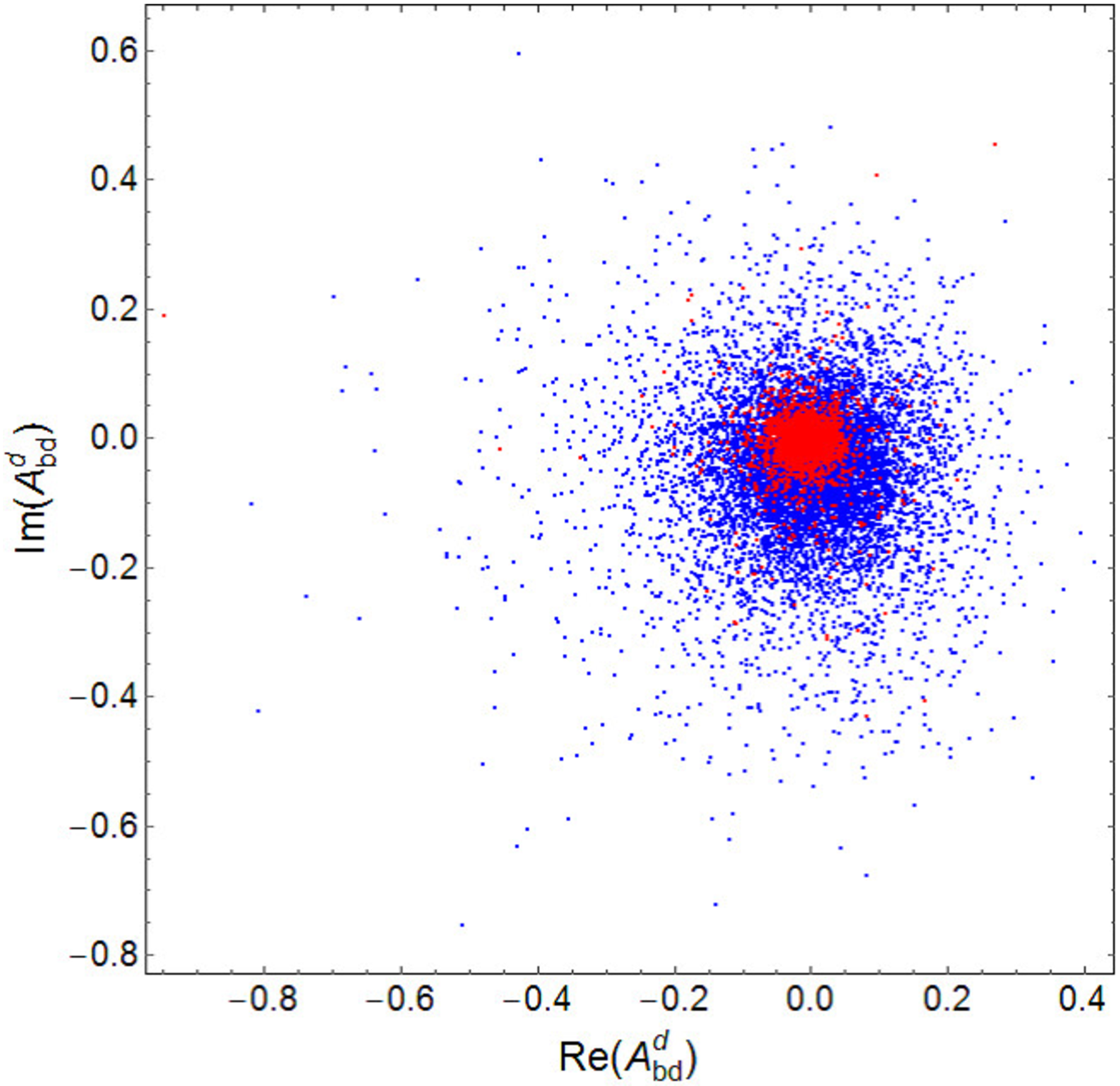,width=0.45\textwidth}}\hspace{0.5cm}{\epsfig{figure=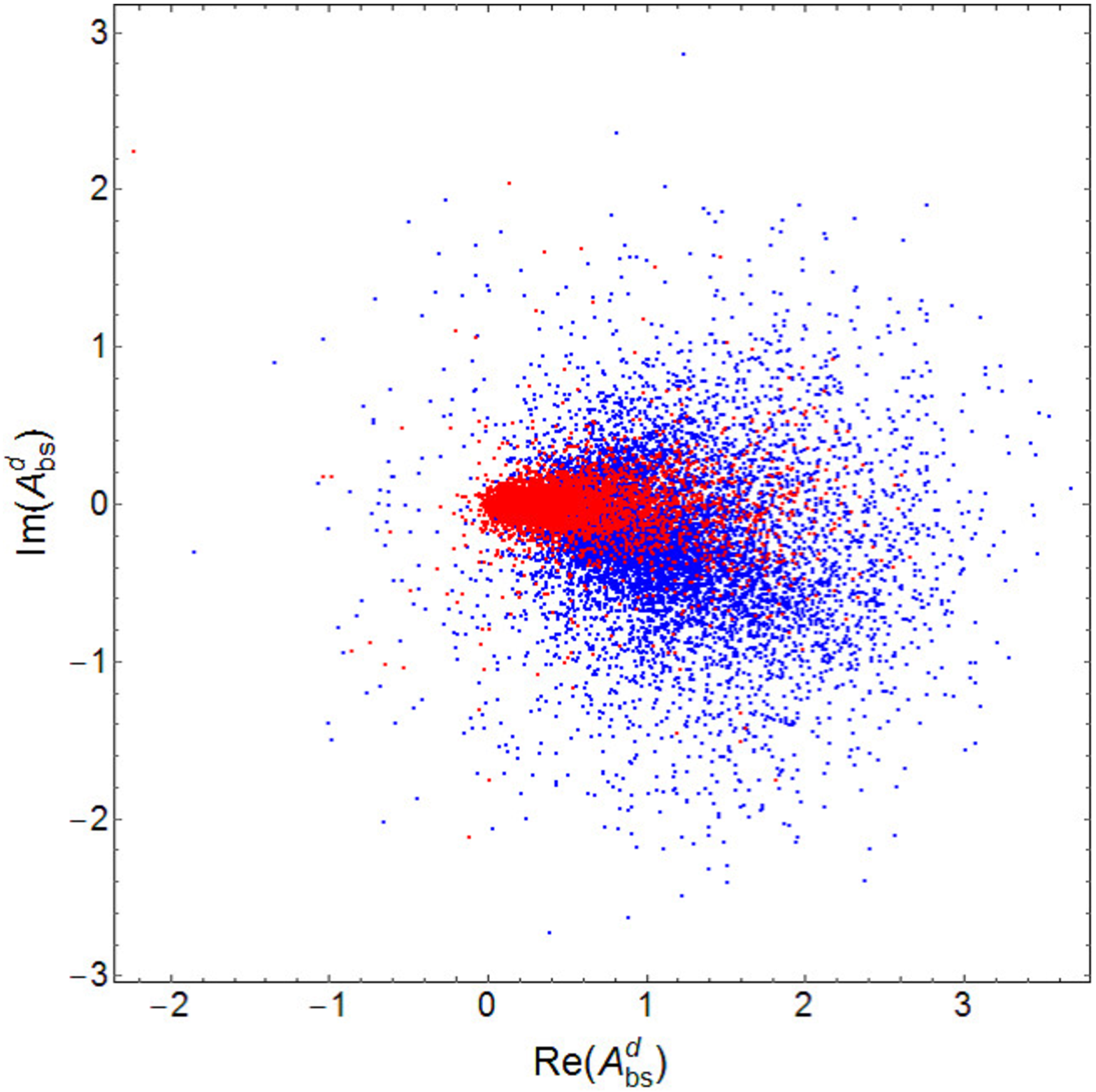,width=0.45\textwidth}}
\caption{ Our predictions for $A^d_{bd}$ (left) and $A^d_{bs}$ (right). The coefficients of higher-dimensional operators satisfy $|\epsilon \,c^d_{ij}|<10^{-2}$ (red) and $|\epsilon \,c^d_{ij}|<10^{-3}$ (blue).}
\label{fig;FCNC2}
\end{center}
\end{figure}

\begin{figure}[!t]
\begin{center}
{\epsfig{figure=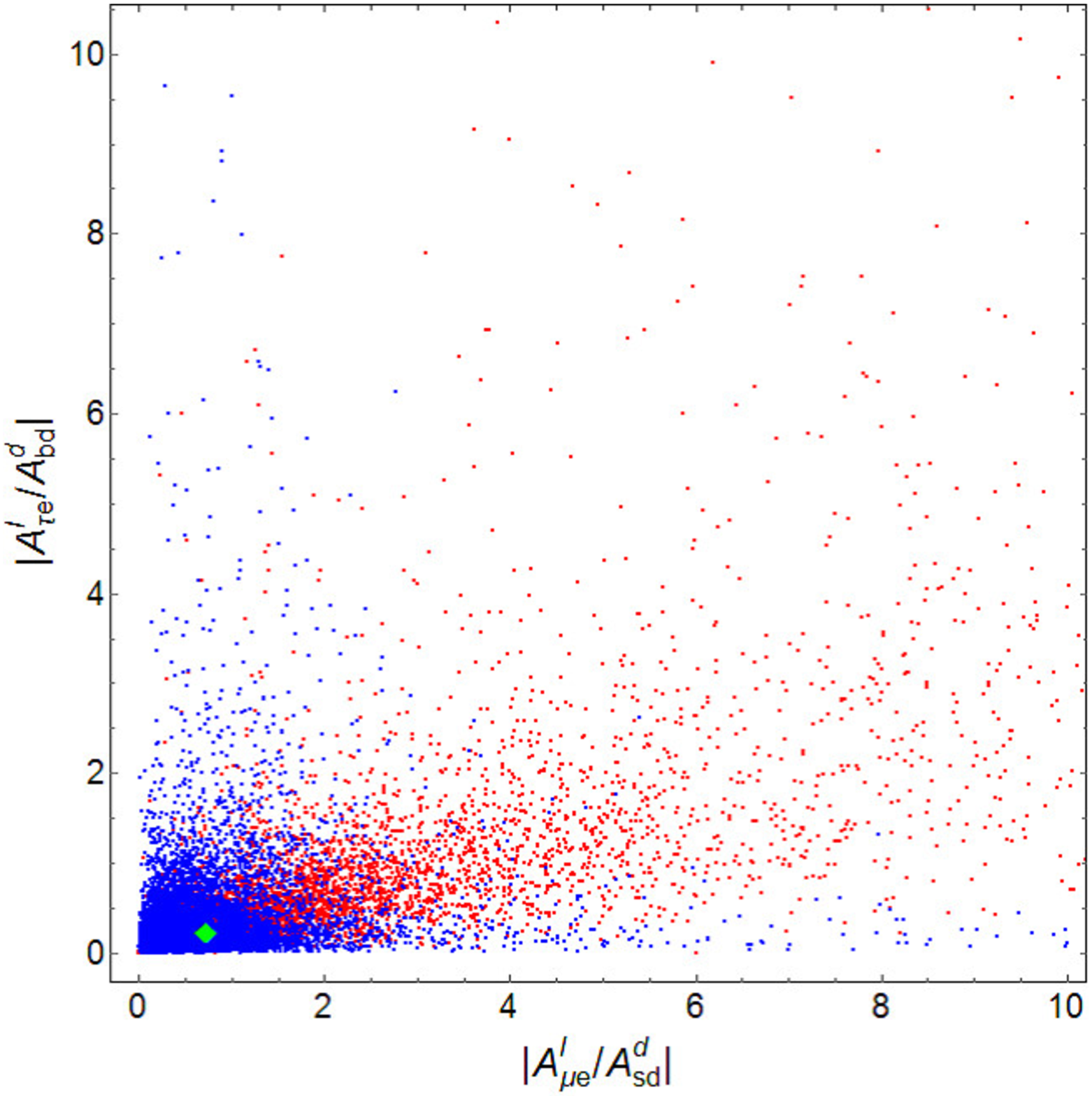,width=0.45\textwidth}}\hspace{0.5cm}{\epsfig{figure=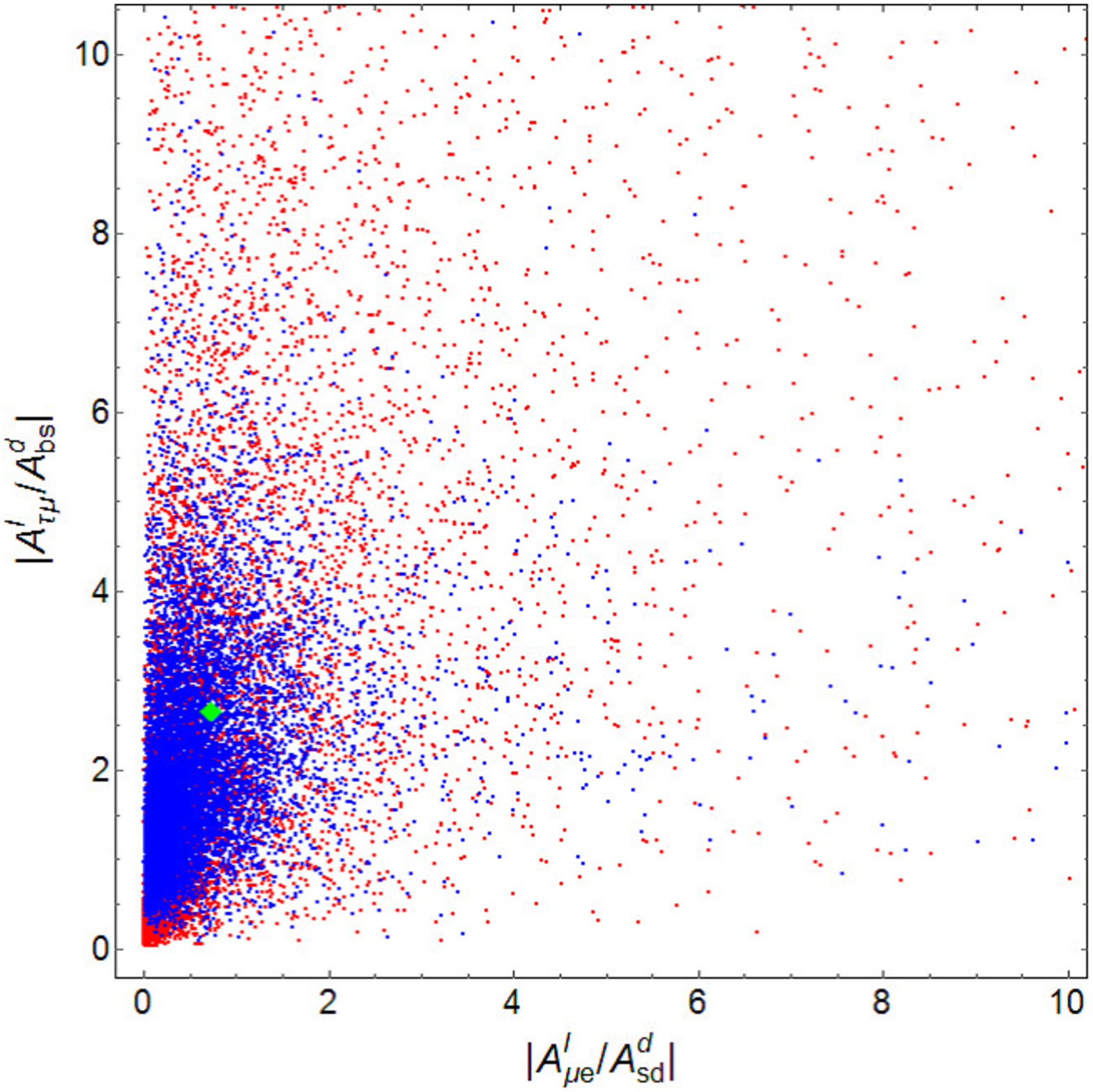,width=0.45\textwidth}}
\caption{ Our predictions for SU(5) relation. The coefficients of higher-dimensional operators satisfy $|\epsilon \, c^{d,l}_{ij}|<10^{-2}$ (red) and $|\epsilon \, c^{d,l}_{ij}|<10^{-3}$ (blue). Green diamond shows the value of each mass ratio of $m^l_im^l_j/(m^d_im^d_j)$.}
\label{fig;SU(5)relation}
\end{center}
\end{figure}

Figs. \ref{fig;FCNC1}, \ref{fig;FCNC2} and \ref{fig;SU(5)relation} show our predictions.
In the all figures of this paper, $\tan \beta$ is fixed at $\tan \beta=3$ and the results in Eqs. (\ref{eq;mass100TeV})
and (\ref{eq;CKM100TeV}) are used. 
In this calculation we assume that $V_R$ is the CKM matrix.
The red (blue) points correspond to arbitral complex $\epsilon \, c^{d,l}_{ij}$ satisfying $|\epsilon \, c^{d,l}_{ij}|<10^{-2}$ ($|\epsilon \, c^{d,l}_{ij}|<10^{-3}$).

Fig. \ref{fig;FCNC1} shows our prediction for $A^{d}_{sd}$ and $A^{d}_{dd}$,
which face the stringent bounds from $K$-$\overline{K}$ mixing.
If we assume the GUT relation in Eq. (\ref{eq;SU(5)relation2}),
those elements are constrained by $\mu \to 3 e$ and $\mu$-$e$ conversion in nuclei as well.
As we see in Fig. \ref{fig;FCNC1}, large $A^{d}_{sd}$ is predicted, so we carefully study the
$K$ physics and $\mu$ physics in Sec. \ref{sec3}.

%%%%%%%%%%%%%%%%%%%%%%%%%%%%%%%%%%%%%%%%%%%%%%%%%%%
%%%%%%%%%%%%%%%%%%%%%%%%%%%%%%%%%%%%%%%%%%%%%%%%%%%
%%%%%%%%%%%%%%%%%%%%%%%%%%%%%%%%%%%%%%%%%%%%%%%%%%%

Let us comment on the mixing to realize the realistic Yukawa coupling.
In the left panel of Fig. \ref{fig;FCNC1}, $A^d_{dd}$ is approximately estimated as $A^d_{dd}\simeq -2$, i.e. $(\Hat U^{d }_{16} )_{1k} (\Hat U^{d }_{16} )^*_{1k} \ll 1$. This means that the SM down quark mainly comes from the {\bf 10}-representational fields of SO(10). The reason is as follows.
We have introduced the higher dimensional operators, suppressed by $\epsilon$,
in order to compensate the small up quark mass. In fact, the contribution to the $(1, \, 1)$ element of the up-type quark mass matrix, denoted by $v_u \epsilon \, c^d_{11} $, is larger than the up quark mass. Then, the down quark mass is roughy given by the suppressed $(\Hat U^{d }_{16} )_{11}$ according to Eq. (\ref{eq;mixingandhighdimd}).

%%%%%%%%%%%%%%%%%%%%%%%%%%%%%%%%%%%%%%%%%%%%%%%%%%%
%%%%%%%%%%%%%%%%%%%%%%%%%%%%%%%%%%%%%%%%%%%%%%%%%%%
%%%%%%%%%%%%%%%%%%%%%%%%%%%%%%%%%%%%%%%%%%%%%%%%%%%

%\begin{figure}[!t]
%\begin{center}
%{\epsfig{figure=Abd.eps,width=0.45\textwidth}}\hspace{0.5cm}{\epsfig{figure=Abs.eps,width=0.45\textwidth}}
%\caption{ Our predictions for $A^d_{bd}$ (left) and $A^d_{bs}$ (right). The coefficients of higher-dimensional operators satisfy $|\epsilon \,c^d_{ij}|<10^{-2}$ (red) and $|\epsilon \,c^d_{ij}|<10^{-3}$ (blue).}
%\label{fig;FCNC2}
%\end{center}
%\end{figure}

On the other hand, it seems that ${\bf 10}$- and ${\bf 16}$-representational fields mix with each other in the second and third generations, as in Figs. \ref{fig;FCNC1} and \ref{fig;FCNC2}. $A^{d}_{sd}$ is relatively smaller than the other off-diagonal elements, but could be ${\cal O}(0.1)$ according to the sizable $(\Hat U^{d }_{16} )_{ij}$. 
We find that $A^{d}_{bs}$ tend to be larger than the other FCNC couplings, in Figs. \ref{fig;FCNC1} and \ref{fig;FCNC2}.
This is because $A^{d}_{ij}$ is proportional to the down-type quark masses, $m^d_i$ and $m^d_j$ $(i,j=d,s,b)$, so roughly speaking, the ratios of $|A^{d}_{bs}/A^{d}_{sd}|$ and $|A^{d}_{bs}/A^{d}_{bd}|$ are $\mathcal{O}(m^d_b/m^d_d)$ and $\mathcal{O}(m^d_s/m^d_d)$, respectively, although the dependences of the quark masses and the CKM elements on $|A^{d}_{ij}|$ are not so simple. When $\epsilon$ is small, the approximate expressions for the flavor violating couplings are
\begin{eqnarray}
&&\hspace{-2.5em}{\rm Re}(A^d_{sd})\sim 5 \tan^2\!\beta  \frac{ m^d_d \, m^d_s}{\left|v_u \epsilon \, c^d_{11}\right|^2}\,\lambda, \:\: {\rm Im}(A^d_{sd})\sim 5 \tan^2\!\beta  \frac{ m^d_d \, m^d_s}{\left|v_u \epsilon \, c^d_{11}\right|^2}\,{\rm Im}\left(\frac{v_u \epsilon \, c^{d\ast}_{12}}{m^u_c}\right), \nonumber \\[1.5ex]
&&\hspace{-0.5em}A^d_{bd}\sim 5 \tan^2\!\beta  \frac{ m^d_d \, m^d_b}{\left|v_u \epsilon \, c^d_{11}\right|^2} \left(\frac{v_u \epsilon \, c^{d\ast}_{12}}{m^u_c}\right) A \lambda^2, \nonumber \\[1.5ex]
&&\hspace{-2.5em}{\rm Re}(A^d_{bs})\sim 5 \tan^2\!\beta  \frac{ m^d_s \, m^d_b}{(m^u_c)^2}\,\lambda^2, \:\: {\rm Im}(A^d_{bs})\sim 5 \tan^2\!\beta  \frac{m^d_s \, m^d_b}{\left|v_u \epsilon \, c^d_{11}\right|^2}\,{\rm Im}\left(\frac{v_u \epsilon \, c^{d\ast}_{12}}{m^u_c}\right) A \lambda^3.
\label{eq;roughAsd}
\end{eqnarray}

These properties are the same for $A^{l}_{ij}$ and then we expect that the ratio between $|A^{d}_{ij}|$ and $|A^{l}_{ij}|$ is predictive even if Eq. (\ref{eq;SU(5)relation}) is failed. When $\epsilon$ is small, the ratio is expected to be $\mathcal{O}(m^l_im^l_j/(m^d_im^d_j))$. Our prediction of the ratio is shown in Fig. \ref{fig;SU(5)relation}. These figures show that these ratios tend to be close to the green diamond, which satisfies $|A^{l}_{ij}/A^{d}_{ij}|=m^l_im^l_j/(m^d_im^d_j)$, in the case with small $\epsilon$. Especially, the convergence is remarkable in the $(2,\, 1)$ elements, $|A^{l}_{\mu e}/A^{d}_{sd}|$.

In addition, $\Hat{Z}'_{\mu}$ is the U(1)$^\prime$ gauge boson, but not the mass eigenstate because of mass mixing between $\Hat{Z}'_{\mu}$ and $Z$ boson denoted by $\Hat{Z}_{\mu}$.
The mass mixing is generated by the U(1)$^\prime$-charged Higgs doublets \cite{Hisano-so10}:
\beq
\begin{pmatrix} \Hat{Z}_{\mu} \\ \Hat{Z}'_{\mu}   \end{pmatrix}=  \begin{pmatrix} \cos \theta & - \sin \theta  \\ \sin \theta & \cos \theta   \end{pmatrix} \begin{pmatrix} Z_{\mu} \\ Z'_{\mu}   \end{pmatrix},
\eeq
where $\sin \theta$ is approximately estimated as
\beq
\label{eq;ZZpmixing}
\tan 2 \theta \simeq 4 \, \frac{g'}{g_Z} \frac{M^2_{Z}}{ M^2_{Z'}}.
\eeq
We have to include this effect, when we discuss the phenomenology in our model.

Note that a scalar from $\Phi$ also has flavor changing Yukawa couplings with the SM fermions and the heavy extra fermions, but the left-handed down-type quarks (right-handed lepton) can be indentified as the heavy fermions, because of the EW symmetry. Then, the flavor violating processes involving the scalar are loop-suppressed and negligibly small in our scenario.

\section{Flavor Physics}
\label{sec3}
In this section, we investigate the flavor violating signals predicted by our SO(10) GUT, based on the setup
discussed above. One of the important predictions is that there are tree-level FCNCs involving $Z'_{\mu}$ and $Z_{\mu}$. Moreover, all elements of the FCNCs could be ${\cal O} (1)$,
corresponding to the higher-dimensional operators. This means that we have to seriously check the consistency with the flavor violating processes concerned with the first and second generations, such as $K$-$\overline{K}$ mixing and $\mu \to 3e$, because the processes are the most sensitive to the new physics contributions. Besides, we find that $(b,\,s)$ element of the $Z'$ coupling becomes larger than the other, so that we investigate the impact of our model on $B$ physics, as well.

%{\bf
The SUSY particle contributions to FCNCs are suppressed by loop factors due to the $R$ parity. However, when the flavor violation in squark mass terms is maximal, the SUSY contributions to the $K$ system may not be negligible even if squark masses are $O(100)$ TeV. Then, we ignore them for simplicity.
%}

First, we study the constraints from the $\Delta F=2$ processes in $K$ and $B_{(s)}$ systems in the next subsection, and then let us discuss the consistency of our model with the observations of the LFV $\mu$ decays in Sec. \ref{sec3-2}. 
We also study the $\Delta F=1$ processes, although the constraints are mild.

%%%%%%%%%%%%%%%%%%%%%%%%%%%%%%%%
%%%%%%%%Input parameters%%%%%%%%%%%%%%%%
\begin{table}
\begin{center}
  \begin{tabular}{|c|c||c|c|} \hline
    $m_K$ & 497.611(13) MeV  \cite{PDG}  & $m_{B_s}$ & 5.3663(6) GeV \cite{PDG}  \\ 
     $F_K$ & 156.1(11) MeV \cite{Lattice1} & $m_{B}$ & 5.2795(3) GeV \cite{PDG} \\ 
     $\Hat B_K$ & 0.764(10) \cite{Lattice1}  &  $F_{B_s}$ & 227.7 $\pm$ 6.2 MeV \cite{Lattice1} \\ 
  $(\Delta M_K)_{\rm exp}$  & 3.484(6)$\times 10^{-12}$ MeV  \cite{PDG} & $F_{B}$ &190.6 $\pm$ 4.6 MeV \cite{Lattice1}   \\ 
    $|\epsilon_K|$ & $(2.228(11)) \times 10^{-3}$  \cite{PDG}  &  $\Hat B_{B_s}$ & 1.33(6) \cite{Lattice1} \\ 
    BR$(K^+ \to \pi^0  \,  e^+  \,  \nu)$ & $5.07(4)$ \%  \cite{PDG} & $\Hat B_{B}$ & 1.26(11)  \cite{Lattice1}  \\ 
         $\tau (K^+)$   & (1.238(2))$\times 10^{-8}$ s  \cite{PDG} & $\eta_B$ & 0.55 \cite{Buras:1990fn}  \\ 
 $\tau (K_L)$ & (5.116(21))$\times 10^{-8}$ s  \cite{PDG}  & $\eta_Y$ &1.012 \cite{Buchalla:1998ba} \\ 
%    $Re(\epsilon'/\epsilon)$& $(1.66(23)) \times 10^{-3}$  \cite{PDG}  \\ \hline
    $\eta_1$ & 1.87(76) \cite{Brod:2011ty} &  $\Gamma^{-1}_\mu$ & $2.1969811(22) \times10^{-6}$ s \\ 
         $\eta_2$& 0.5765(65) \cite{Buras:1990fn} & & \\ 
     $\eta_3$ & 0.496(47) \cite{Brod:2010mj} & & \\ \hline
  \end{tabular}
 \caption{The input parameters relevant to our analyses in flavor physics.}
  \label{table;input2}
  \end{center}
\end{table}
%%%%%%%%%%%%%%%%%%%%%%%%%%%%%%%%
%%%%%%%%%%%%%%%%%%%%%%%%%%%%%%%%

\subsection{$\Delta F=2$ processes }
In the SM, CP violation is caused by the CP phase in the CKM matrix. 
CP violating processes as well as flavor violating processes are strongly suppressed by the GIM mechanism,
and the SM predictions are usually very tiny. The flavor processes of $K$ meson are no exception.
In fact, the SM prediction of $K$-$\overline{K}$ mixing is quite small, but it is consistent with the experimental observations, although there are still sizable theoretical uncertainties in the SM predictions.
In other words, large new physics contributions to the $K$ physics conflict with the experimental results, and then the strong constraints should be taken into account.
Similarly, we can derive the new physics constraints from $B$-$\overline{B}$ and $B_s$-$\overline{B_s}$ mixing.

In addition to the SM corrections, the $\Delta F=2$ processes are caused by the tree-level FCNCs of $Z'$ and $Z$ in our model. The induced operators are 
\begin{equation}
{\cal H}^{\Delta F=2}= \frac{1}{2} \sum_{q=K,B,B_s}\widetilde{ C}^{q}_1 \widetilde Q^q_1 
\end{equation}
where the each operator is given by
\beq
\widetilde Q^K_1=  (\overline{s_R} \gamma_\mu d_R)(\overline{s_R} \gamma^\mu d_R),~
\widetilde Q^{B}_1=  (\overline{b_R} \gamma_\mu d_R)(\overline{b_R} \gamma^\mu d_R),~
\widetilde Q^{B_s}_1=  (\overline{b_R} \gamma_\mu s_R)(\overline{b_R} \gamma^\mu s_R),
\eeq
and the Wilson coefficient is estimated as
\beq
\label{eq;C1}
\widetilde{ C}^K_1 =  (A^d_{sd})^2 \,\left( \frac{g'^2\cos^2 \theta}{M^2_{Z'}} +\frac{g'^2 \sin ^2 \theta}{M^2_{Z}}  \right ) \equiv  \frac{(A^d_{sd})^2}{\Lambda_{Z'}^2} .
\eeq
$ \widetilde{ C}^{B}_1$ and $\widetilde{ C}^{B_s}_1$ can be derived by exchanging $(A^d_{sd})$ in $\widetilde{ C}^K_1$ with $(A^d_{bd})$ and $(A^d_{bs})$ respectively.
Note that the SM correction appears in the $C^q_1$ ($q=K, \, B, \, B_s$), which are the coefficients of the operators that consist of only left-handed quarks, instead of the right-handed in $\widetilde Q^q_1$: for example,
$(\overline{s_L} \gamma_\mu d_L)(\overline{s_L} \gamma^\mu d_L)$. The CP-phase appears in the $(t, \,d)$-element of the CKM matrix in the SM.
In our model, the FCNCs, $A^d_{ij}$, are generally complex, so that the CP-violating processes strongly constrain our $Z'$ interaction. 

In our analyses on flavor physics, we set $\Lambda_{Z'} = 1.4 \times 10^{3}$ TeV (500 TeV), which corresponds to $M_{Z'}=100$ TeV (36 TeV) and $g' \simeq 0.073$ \cite{Hisano-so10}. 
$\tan \beta$ is fixed at $\tan \beta=3$ to achieve the 125 GeV Higgs mass \cite{HSSUSY}. 
%The red (blue) points correspond to arbitral complex $\epsilon \, c^{d,l}_{ij}$ satisfying $|\epsilon \, c^{d,l}_{ij}|<10^{-2}$ ($|\epsilon \, c^{d,l}_{ij}|<10^{-3}$).

\subsubsection{$\Delta S =2$ process}

Based on Ref. \cite{Buras:2012jb}, we investigate the upper bound on the $Z'$ interaction from the $K$-$\overline{K}$ mixing.
The physical observables on the mixing are denoted by $\epsilon_K$ and $\Delta M_K$,
which are evaluated as 
\beq
\epsilon_K= \frac{\kappa_\epsilon e^{i \varphi_\epsilon} }{\sqrt{2} (\Delta M_K)_{\rm exp}} \, {\rm Im}(M^K_{12}), ~ \Delta M_K =2  {\rm Re}(M^K_{12}).
\eeq
$\kappa_\epsilon$ and $\varphi_\epsilon$ are given by the observations:
$\kappa_\epsilon=0.94 \pm 0.02$ and $\varphi_\epsilon=0.2417 \times \pi$.
$M^K_{12}$ is generated by the $K$-$\overline{K}$ mixing and decomposed as follows in our model:
\beq
M^K_{12}=\left ( M^K_{12} \right )_{\rm SM} + \Delta M^K_{12}.
\eeq
$\Delta M^K_{12}$ is the $Z'$ contribution, and then it is given by
\beq
\Delta M^K_{12}= \frac{1}{2} \, \widetilde{ C}^K_1(\mu) \langle \widetilde Q^K_1 \rangle. 
\eeq
The matrix element, $ \langle \widetilde Q^K_1 \rangle$, can be extracted from the SM prediction, because the only difference is the chirality. $ \widetilde{ C}^K_1(\mu)$ is the Wilson coefficient derived from Eq. (\ref{eq;C1}) and the RG correction. The running correction is studied in Appendix \ref{sec;appendix1-2}.

The SM prediction is described as
\beq
(M^K_{12})_{\rm SM}= \frac{G^2_F}{12 \pi^2} F^2_K \Hat{B}_K m_K M^2_W \left \{ \lambda^2_c \eta_1S_0( x_c) +  \lambda^2_t \eta_2 S_0(x_t) + 2  \lambda_c \lambda_t \eta_3 S(x_c, x_t)  \right \}.
\eeq
$x_i$ and $\lambda_i$ denote $m^2_i/M^2_W$ and $(V_{CKM})^*_{is} (V_{CKM})_{id}$, respectively. $\eta_{1,2,3}$ correspond to the NLO and NNLO QCD corrections \cite{Brod:2011ty,Buras:1990fn,Brod:2010mj}. The values we adopt are summarized in Table \ref{table;input2}. 
The functions,  $S_0(x_t)$ and $S(x_c, x_t)$, are shown in Appendix \ref{sec;appendix1}.

The physical observables in $K$-$\ov{K}$ mixing are experimentally measured well.
On the other hand, the SM predictions still suffer from the large uncertainty from the matrix element and the CKM matrix. Using the central values in Table \ref{table;input} and Table \ref{table;input2}, we draw our predictions for the deviations of $\epsilon_K$ and $\Delta M_K$ from the SM predictions.
%-----------------
\begin{figure}[!t]
\begin{center}
{\epsfig{figure=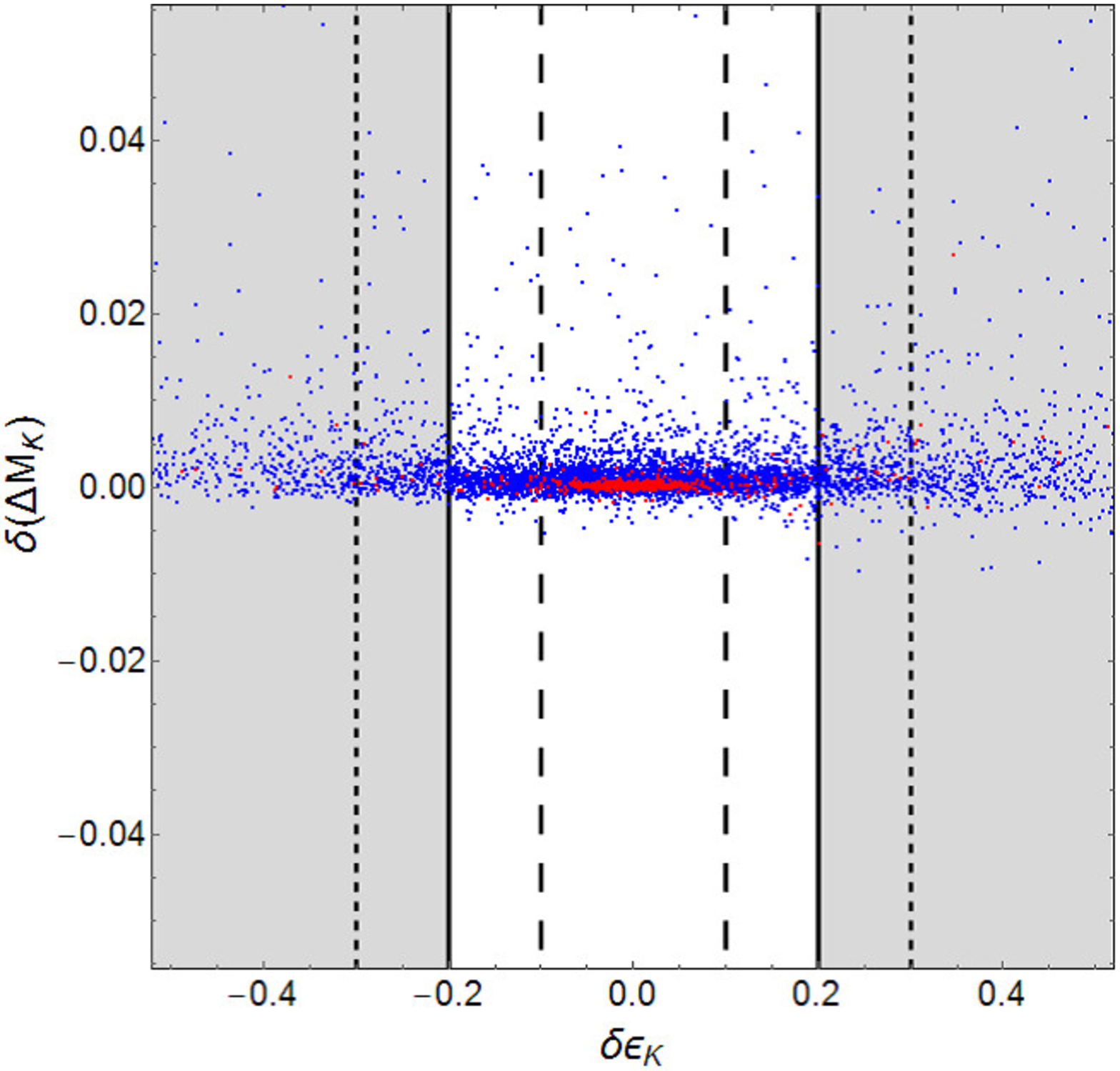,width=0.45\textwidth}}\hspace{0.5cm}{\epsfig{figure=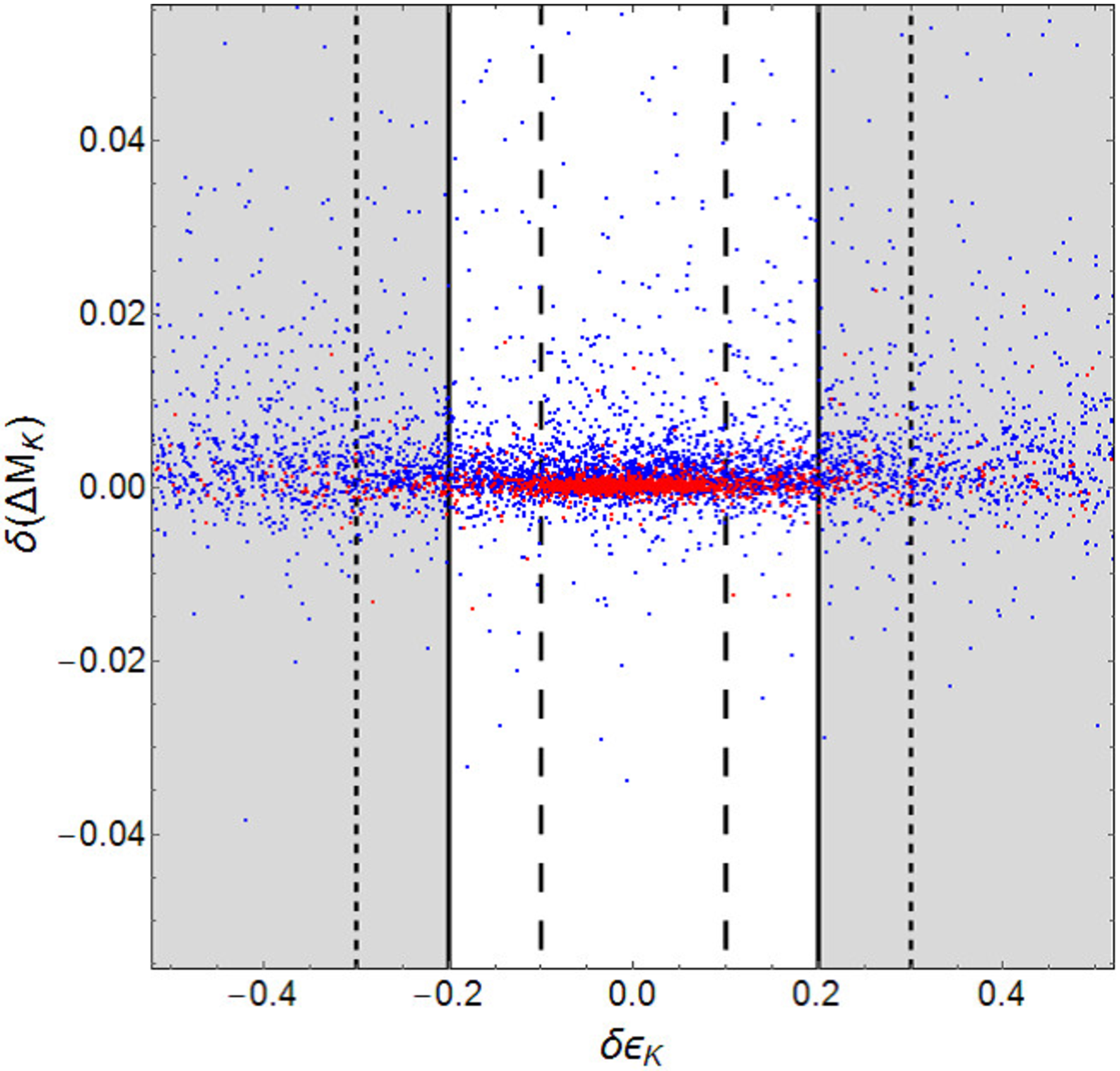,width=0.45\textwidth}}
\caption{ Our predictions for $\delta \epsilon_K$ and $\delta (\Delta M_K)$ with $\Lambda_{Z'}=1400$ TeV (left) and $\Lambda_{Z'}=500$ TeV (right). The coefficients of higher-dimensional operators satisfy $|\epsilon \,c^d_{ij}|<10^{-2}$ (red) and $|\epsilon \,c^d_{ij}|<10^{-3}$ (blue). Black dashed, solid and dotted line show the deviation from SM by $10\%$, $20\%$ and $30\%$, respectively.}
\label{fig;KKbarmixing}
\end{center}
\end{figure}
%------------------------------
Compared to the SM predictions, $(\epsilon_K)_{\rm SM}$ and $(\Delta M_K)_{\rm SM}$,
the deviations are defined as 
\beq
\delta \epsilon_K \equiv \epsilon_K/  (\epsilon_K)_{\rm SM}-1~~{\rm and}~ ~\delta (\Delta M_K) \equiv \Delta M_K/  (\Delta M_K)_{\rm SM}-1.
\label{eq;deviations}
\eeq  
It is difficult to draw the exclusion limits in terms of $|\delta \epsilon_K|$ and $| \delta (\Delta M_K)|$,
because of the large uncertainties of the SM predictions.
In Ref. \cite{Charles:2013aka}, the CKM fitter group shows that the experimental upper bounds on  $|\delta \epsilon_K|$ and $| \delta (\Delta M_K)|$ are at most O(30) \%. It will be developed up to O(20) \% at the Belle II experiment  \cite{Charles:2013aka}.

In Fig. \ref{fig;KKbarmixing}, our predictions for the deviations of $\epsilon_K$ and $\Delta M_K$ are shown
in the cases with $\Lambda_{Z'}=1400$ TeV (left) and $\Lambda_{Z'}=500$ TeV (right).
The black dashed, solid and dotted line show the deviation from SM by $10\%$, $20\%$ and $30\%$, respectively.
In our model, $\epsilon_K$ largely departs from the SM prediction, even if $\Lambda_{Z'}$ is ${\cal O}(10^3)$ TeV.
Then, we have to consider the consistency with $\epsilon_K$, whenever we discuss the other observables.
%------------------------------------------------------------------------------
\subsubsection{$\Delta B=2$ process}
We now derive our predictions of $B$-$\overline{B}$ and $B_s$-$\overline{B_s}$ mixing,
as well as $K$-$\overline{K}$ mixing. The observables relevant to the mixing 
are mass differences denoted by $\Delta M_{B}$ and $\Delta M_{B_s}$.
They are influenced by $\widetilde C^{B}_1$ and $\widetilde C^{B_s}_1$ as follows:
\beq
\Delta M_{B_q}=  2 \left | (M^{B_q}_{12})_{\rm SM} + \frac{1}{6} \widetilde C^{B_q}_1 m_{B_q} F_{B_q} \Hat B_{B_q} \right | ~(q=d, \, s),
\eeq
where $ (M^{B_q}_{12})_{\rm SM}$ is given by the top-loop contribution:
\beq
 (M^{B_q}_{12})_{\rm SM}=\frac{G^2_F}{12 \pi^2} F^2_{B_q} \Hat{B}_{B_q} m_{B_q} M^2_W   \lambda^2_{B_q } \eta_BS_0(x_t) .
\eeq
The input parameters used in our analyses are shown in Table \ref{table;input2}. $\lambda_{B_q }$ depicts  
$\lambda_{B_q }=(V_{CKM})^*_{tb} (V_{CKM})_{tq}$. 
The SM predictions still have large uncertainties dominated by the errors of hadronic mixing matrix elements and the CKM matrix elements, so that it would be difficult to draw the new physics constraints as well. 
Recently, the Fermilab and MILC Collaborations have shown their results on the SM predictions of  $\Delta M_{B}$ and $\Delta M_{B_s}$ \cite{Bazavov:2016nty} and about 10 \% errors are still inevitable.
The LHCb and Belle II experiments will improve the measurement, as discussed in Ref. \cite{Charles:2013aka}.

\begin{figure}[!t]
\begin{center}
{\epsfig{figure=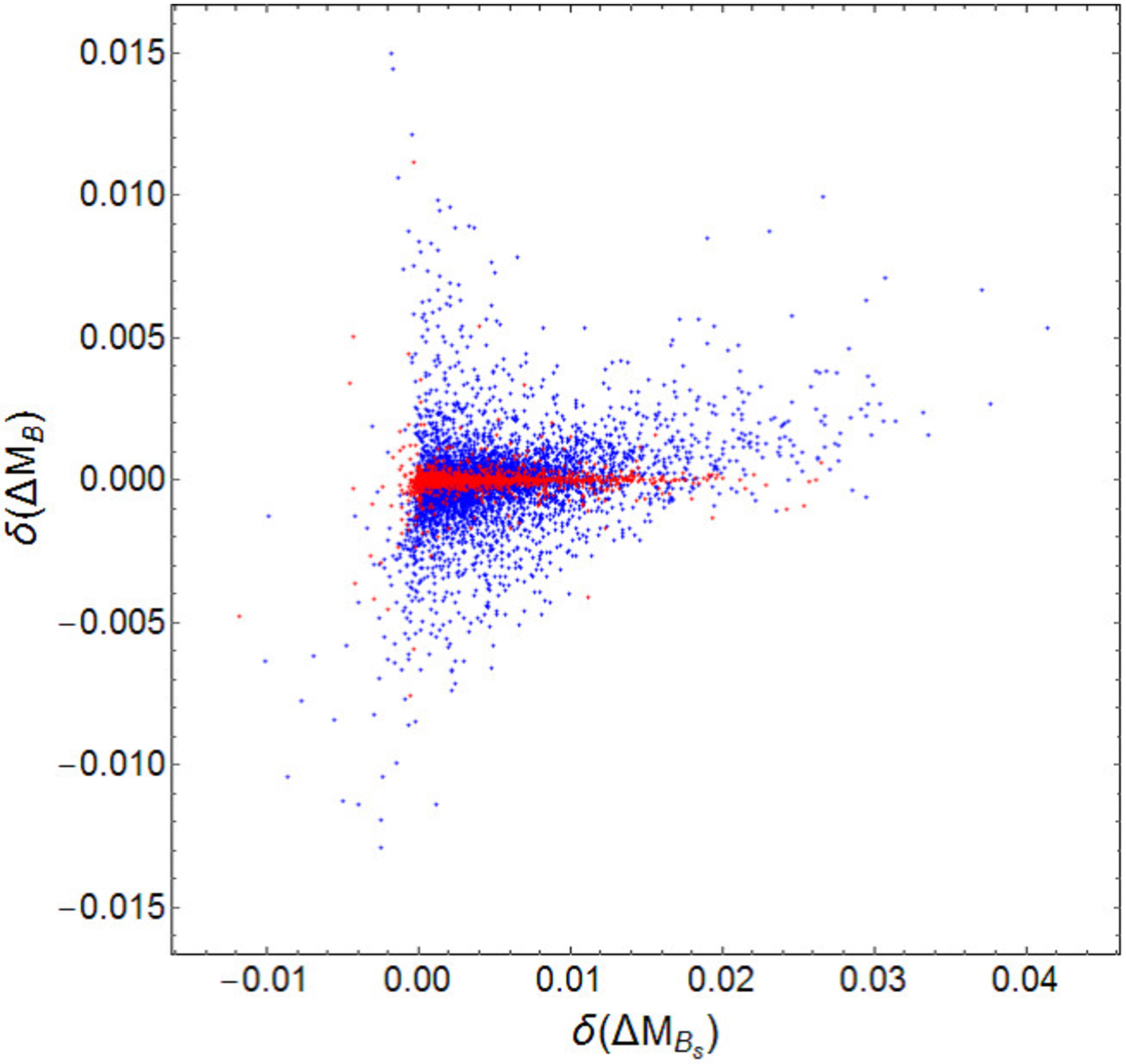,width=0.45\textwidth}}\hspace{0.5cm}{\epsfig{figure=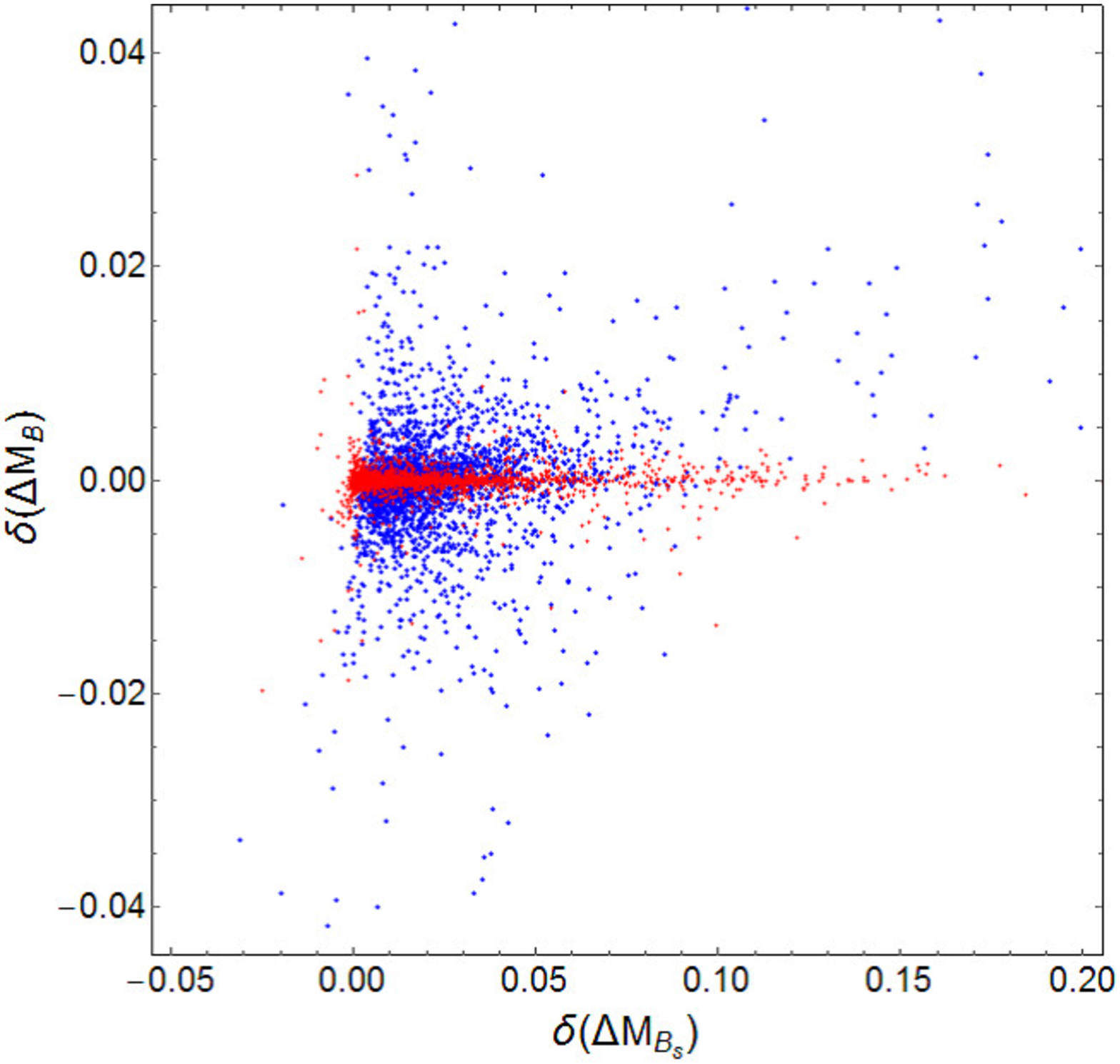,width=0.45\textwidth}}
\caption{ Our predictions for $\delta (\Delta M_{B_s})$ and $\delta (\Delta M_B)$ with $\Lambda_{Z'}=1400$ TeV (left) and $\Lambda_{Z'}=500$ TeV (right). The coefficients of higher-dimensional operators satisfy $|\epsilon \,c^d_{ij}|<10^{-2}$ (red) and $|\epsilon \,c^d_{ij}|<10^{-3}$ (blue). 
In these figures, we only show the points that $|\delta \epsilon_K| \leq 0.3$ is satified.}
\label{fig;BBbarmixing}
\end{center}
\end{figure}

In our model, $A^d_{bs}$ is large compared to the other elements, so that our model may be tested by $\Delta M_{B_s}$, although the deviation is relatively smaller than the $K$-$\overline{K}$ mixing because of the size of the SM prediction. 
Fig. \ref{fig;BBbarmixing} shows our predictions for the deviations of $\Delta M_{B_s}$ and $\Delta M_B$
in the cases with $\Lambda_{Z'}=1400$ TeV (left) and $\Lambda_{Z'}=500$ TeV (right). 
These deviation are defined as the same manner in Eq. (\ref{eq;deviations}). 
If $Z'$ is around ${\cal O}(10)$ TeV, $\delta(\Delta M_B)$ could reach 10 \%, which maybe cause the tension with
the current measurement \cite{Charles:2013aka}. In these figures, all points satisfy $|\delta \epsilon_K| \leq 0.3$.

%%%%%%%%%%%%%%%%%%%%%%%

\subsection{$\Delta F =1$ processes}
%%%%%%%%%%%%%%%%%%%%%%%
The $Z'$ interaction deviates the SM predictions in the rare decays of $B$ and $K$ mesons.
The KOTO, Belle II and LHCb experiments will develop the measurements of the rare decays and give some hints to new physics.
In this section, we study the (semi) leptonic decays of $K$ meson and the leptonic decays of $B$ and $B_s$.
The processes we especially study here are $K_L \to \pi^0 \nu \overline{\nu} $, measured by the KOTO experiment, 
$K_L \to \mu^+ \mu^-$, $\mu^{\pm} e^{\mp}$ and $B_s (B) \to \mu^+ \mu^-$.

\subsubsection{$\Delta S =1$ processes}
The $\Delta S =1$ processes, such as the rare $K$ meson decays, play a crucial role in testing our model. The effective Hamiltonian which causes the tree-level flavor changing is given by
the $Z'$ exchanging and $Z$ boson exchanging through the $Z$-$Z'$ mixing:
\beq
\label{eq;deltaS1}
{\cal H}^{\Delta S=1} =(C^f_I)_{ij} (\overline{s_R} \gamma_\mu d_R)(\overline{f^i_I} \gamma^\mu f^j_I),
\eeq
 where $f$ denotes $f=\nu,\, l, \, u, \, d$ and $I$ is the chirality of the fermions ($f$) ($I=L, \, R$).
 $(C^f_I)^{ij}$ at $\mu=M_{Z'}$ is described as
 \beq
 \label{eq;deltaS1-1}
 (C^f_I)_{ij} =-A^d_{sd}  \left \{ \frac{ (Q^f_{I})_{ij} }{ \Lambda^2_{Z'} }+  \frac{ \delta_{ij} }{\Lambda^2_{Z} } \left ( \tau^f_I - Q_e^{f} \sin^2 \theta_W \right )  \right \},
 \eeq
 where $\Lambda^2_{Z}$ is defined as
 \beq
 \frac{1}{ \Lambda^2_{Z}}= g' g_Z \sin \theta \cos \theta  \left ( \frac{1}{M^2_{Z}}-\frac{1}{M^2_{Z'}} \right ).
 \eeq
  Note that the second term  $\Lambda^2_{Z}$ is approximately evaluated as $\Lambda^2_{Z} \simeq \Lambda^2_{Z'}/2 $ according to Eq.(\ref{eq;ZZpmixing}), in the limit $M_{Z'} \gg M_Z$.
 $ (Q^f_{I})_{ij}$ are given by Eq. (\ref{eq;gauge}) as
 \begin{eqnarray}
 ((Q^{\nu,l}_{L})_{ij}, \, (Q^l_{R})_{ij})&=& ( A^l_{ij}, \, +\delta_{ij}),  \\
  ((Q^{u,d}_{L})_{ij}, \, (Q^u_{R})_{ij} \, , \, (Q^d_{R})_{ij})&=&  ( -\delta_{ij}, \, +\delta_{ij}, -A^d_{ij}). 
 \end{eqnarray} 
 $\tau^f_I$ and $Q_e^{f}$ are the isospin and the EW charge of $f$.
 In this subsection, we study the impacts of these new physics corrections on the $K$ meson decays.

\subsubsection*{ \underline{$K_L \to \pi^0  \,  \nu  \,  \overline{\nu}$ and $K^+ \to \pi^+  \,  \nu  \,  \overline{\nu}$ }}
Another important measurement of the CP-violating processes is the rare decay of neutral $K$ meson:
$K_L \to \pi^0 \, \nu \, \overline{\nu}$. The SM prediction is quite tiny, and it is not still reached by the past and current experiments: BR$(K_L \rightarrow \pi^0 \nu \overline{\nu})< 2.6 \times 10^{-8}$ \cite{E391a}. The KOTO experiment at the J-PARC will cover the region near future.
On the other hand, the decay of the charged $K$ meson, $K^+ \to \pi^+ \, \nu \, \overline{\nu}$,
has been already measured as BR($K^+ \rightarrow \pi^+ \nu \overline{\nu}) =1.73^{+1.15}_{-1.05}  \times 10^{-10}$ \cite{E949} and will be updated by the NA62 experiment at the CERN.

In the SM, the both branching ratios are given by the following operators, 
\beq
{\cal H}_{\rm{SM}}^{\Delta S=1} =C_{\rm SM} (\overline{s_L} \gamma_\mu d_L)(\overline{\nu_L}^i \gamma^\mu \nu^i_L).
\eeq
$C_{\rm SM} $ is given by the $Z$ penguin diagram and the box diagram involving $W$ boson.
Then,  the SM prediction, $C_{\rm SM}$, is described in the following form, \footnote{See Ref. \cite{Buras:2015qea} for the current status of the calculations. } 
\beq
C_{\rm SM}=\frac{G_F}{\sqrt{2}} \frac{2 \, \alpha}{ \pi \sin^2 \theta_W} \left ( \lambda_c X_c+ \lambda_t X(x_t) \right  ).
\eeq 
$X_c/ \lambda^4=(0.42 \pm 0.03)$  is proposed in Ref. \cite{Buras:2012jb}.
$X(x_t)$ is the short-distance contribution given by the $Z$-penguin diagrams and box diagrams involving top quark respectively. We can see the LO description in Appendix \ref{sec;appendix1}.
In addition, we have the $Z'$ contribution to this process, as we see in Eqs. (\ref{eq;deltaS1}) and (\ref{eq;deltaS1-1}).
Using the isospin symmetry and taking the ratio to $K^+ \to \pi^0  \,  e^+  \,  \nu$,
the branching ratio of $K_L \to \pi^0  \,  \nu  \,  \overline{\nu}$ is estimated as
 \beq
 \label{eq;KtoPinunu}
 {\rm BR} (K_L \to \pi^0  \,  \nu  \,  \overline{\nu}) = \frac{{\cal A}_{ij} {\cal A}^*_{ij}}{ 8|(V_{CKM})_{us}|^2G^2_F} \times \frac{\tau (K_L)}{\tau (K^+)} \times r_{K_L} \times {\rm BR} (K^+ \to \pi^0  \,  e^+  \,  \nu),
 \eeq
 where ${\cal A}_{ij}$ is defined as
 \beq
 {\cal A}_{ij}=\frac{1}{\sqrt{2}} \left \{ \delta_{ij}  (C_{\rm SM}-C^*_{\rm SM}) + (C^{\nu}_{L})_{ij}-(C^{\nu }_{L})^*_{ji} \right \}.
 \eeq
$r_{K_L}$ is the isospin breaking effect. Based on Ref. \cite{Mescia:2007kn}, we estimate it as $r_{K_L} \simeq 0.955$. Note that the SM prediction is BR$(K_L \to \pi^0 \nu \ov{\nu})=2.43(39)(6) \times 10^{-11}$ \cite{Brod:2010hi}.

 $K_L \to \pi^0  \,  \nu  \,  \overline{\nu}$ is the CP-violating process, so that the decay depends on the imaginary part of the tree-level FCNCs. ${\rm BR} (K^+ \to \pi^0  \,  e^+  \,  \nu)$ is well measured at the experiments, and 
 we can expect that the $Z'$ contribution to $K^+ \to \pi^0  \,  e^+  \,  \nu$ is rather small in this process.
 Then we use the experimental result as the input parameter.
 Note that the penguin diagram contribution to $C_{\rm SM}$ is also modified by $\cos^2 \theta$,
 but here we ignore such a new physics contribution at the one loop level.

 %%%%%%%%%%%%%%%%%%%%%%%%%%%%%%%%%%%%%%%%%

Similarly, we can estimate the branching ratio of $K^+ \to \pi^+  \,  \nu  \,  \overline{\nu}$,
 \beq
 {\rm BR} (K^+ \to \pi^+  \,  \nu  \,  \overline{\nu}) =
\frac{ {\cal A}^+_{ij}  {\cal A}^{+ \, *}_{ij}}{ 8|(V_{CKM})_{us}|^2G^2_F} \times r_{K^+} \times {\rm BR} (K^+ \to \pi^0  \,  e^+  \,  \nu),
 \eeq
 where ${\cal A}^+_{ij}$ is given by
 \beq
 {\cal A}^+_{ij}=\sqrt{2} \left \{ \delta_{ij} C_{\rm SM} + (C^{\nu}_{L})_{ij} \right \}.
 \eeq
 We estimate the isospin breaking effect, $r_{K^+}$, as $r_{K^+} \simeq 0.978$ \cite{Mescia:2007kn}.
  Note that the SM prediction is BR$(K^+ \to \pi^+ \nu \ov{\nu})=7.81(75)(29) \times 10^{-11}$ \cite{Brod:2010hi}.

\begin{figure}[!t]
\begin{center}
{\epsfig{figure=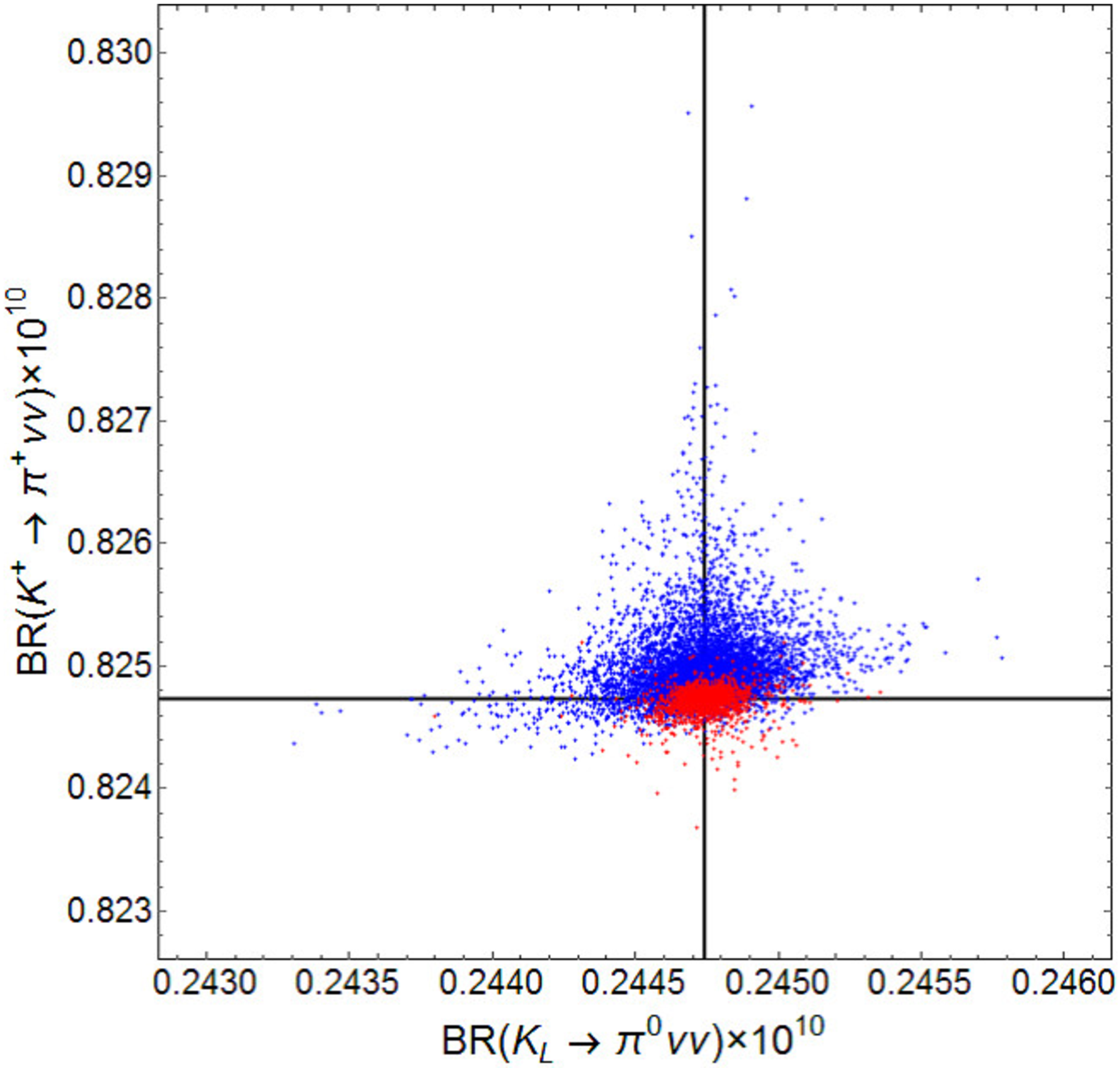,width=0.45\textwidth}}\hspace{0.5cm}{\epsfig{figure=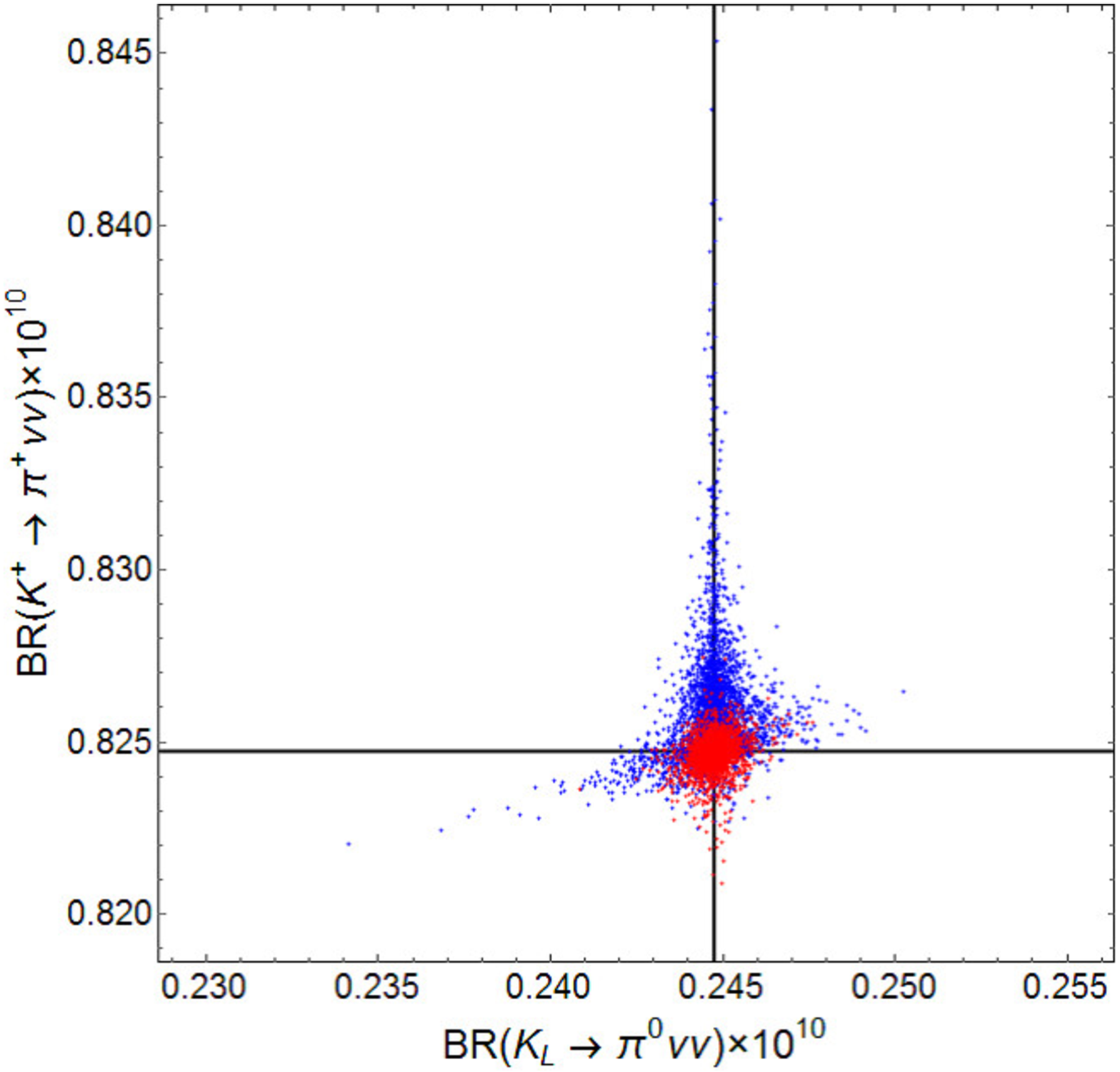,width=0.45\textwidth}}
\caption{ Our predictions for BR($K_L \rightarrow \pi^0 \nu \overline{\nu})$ and BR($K^+ \rightarrow \pi^+ \nu \overline{\nu})$ with $\Lambda_{Z'}=1400$ TeV (left) and $\Lambda_{Z'}=500$ TeV (right). The coefficients of higher-dimensional operators satisfy $|\epsilon \,c^d_{ij}|<10^{-2}$ (red) and $|\epsilon \,c^d_{ij}|<10^{-3}$ (blue). Black solid lines show each SM prediction. 
The all points satisfy $|\delta \epsilon_K| \leq 0.3$.}
\label{fig;Kpinunuzoom}
\end{center}
\end{figure}

Fig. \ref{fig;Kpinunuzoom} shows our predictions of BR($K_L \rightarrow \pi^0 \nu \overline{\nu})$ and BR($K^+ \rightarrow \pi^+ \nu \overline{\nu})$, satisfying $|\delta \epsilon_K| \leq 0.3$.
Black solid lines show the SM predictions, using the center values in Table \ref{table;input2}. 
BR($K^+ \rightarrow \pi^+ \nu \overline{\nu})$ tends to be slightly larger than SM prediction. This is because this deviation is proportional to ${\rm Re}(C_{SM} A^d_{sd}) \sim {\rm Re}(C_{SM}) {\rm Re}(A^d_{sd})$, where Re($A^d_{sd}$) tends to be positive, as shown in Fig. \ref{fig;FCNC1} and Eq. (\ref{eq;roughAsd}). On the other hand, the dominant deviation of BR($K_L \rightarrow \pi^0 \nu \overline{\nu})$ is proportional to ${\rm Im}(C_{SM}) {\rm Im}(A^d_{sd})$. Therefore, such a specific trend can not be seen in BR($K_L \rightarrow \pi^0 \nu \overline{\nu})$.
In any case, our predictions do not largely depart from the SM prediction, as far as $\Lambda_{Z'}=1.4 \times 10^3$ TeV.
Even if $\Lambda_{Z'}$ is around 500 TeV, the deviation is at most 10 \%, compared to the SM prediction.

  %%%%%%%%%%%%%%%%%%%%%%%

 \subsubsection*{ \underline{ $K_L \to l_i \,  l_j$ and $K_L \to \pi^0 \, l_i \,  l_j$}}
 The leptonic decays of $K_L$ may be also important in our model.
 $K_L \to \mu^+ \mu^-$ has a large long-distance contribution in the decay width.
 In Ref. \cite{Ishidori}, the new physics constraint from $K_L \to \mu^+ \mu^-$ 
 is proposed, extracting the  the short-distance part: BR$ (K_L \rightarrow \mu^+ \mu^-) < 2.5 \times 10^{-9}$.
 In our model,  the branching ratio of $K_L \to \mu^+ \mu^-$ departs from the SM prediction because of the flavor changing $Z'$ couplings. The extra contribution is depicted by $(C^l_{L, \,R})_{\mu \mu}$ defined in Eq. (\ref{eq;deltaS1}).
Following Refs. \cite{Buras:2012jb,Buras:2004ub,Gorbahn:2006bm}, we estimate the deviation of this leptonic decay.
As we have already seen above, our prediction cannot be far from the SM one, as far as $\Lambda_{Z'}=\mathcal{O}(10^3)$ TeV. Then, the short-distance part of BR($K_L \to \mu^+ \mu^-$) is dominated by SM contribution, so that the ratio between our prediction and the SM one of BR($K_L \to \mu^+ \mu^-$) is estimated as
\begin{eqnarray}
\left|\frac{BR(K_L \to \mu^+ \mu^-)}{BR(K_L \to \mu^+ \mu^-)_{\rm SM}} - 1\right| \leq 0.019,
\end{eqnarray}
when $\Lambda_{Z'}=1.4 \times 10^3$ TeV.
We conclude that the bound from this process does not threaten our model in the high-scale SUSY scenario. 

The flavor violating decay of $K_L$ has been experimentally investigated as well: $K_L \rightarrow \mu^+ e^- <4.7 \times 10^{-12}$  \cite{BNL-E871}. 
Similarly, BR($K_L \to \mu^+ e^-$) also cannot be large in our model.
Using $\Lambda_{Z'}=1.4 \times 10^3$ TeV and typical values of $A^d_{sd}$ and $A^l_{\mu e}$, this branching ratio is
\begin{eqnarray}
{\rm BR}(K_L \to \mu^+ e^-) \simeq 4.0 \times 10^{-19} \left(\frac{1400\,{\rm TeV}}{\Lambda_{Z'}}\right)^4 \left(\frac{{\rm Re}(A^d_{sd})}{0.1}\right)^2 \left(\frac{|A^l_{\mu e}|}{0.04}\right)^2.
\end{eqnarray}
This is much below the experimental bound.

The semileptonic decay of $K$ such as $K_L \to \pi^0 \, \overline{l}_i \, l_j$ may be relevant to our model.
 The current experimental upper bounds are \cite{KTeVee:2003,KTeVmumu:2000}
\begin{eqnarray}
{\rm BR} (K_L \to \pi^0 e^+ e^-)&<&2.8 \times 10^{-10},  \\ 
{\rm BR} (K_L \to \pi^0 \mu^+ \mu^-)&<&3.8 \times 10^{-10},
\end{eqnarray}
which are about 10 times bigger than the SM predictions  \cite{Mescia:2006jd}, so that large new physics effects are still allowed in these decay modes. Similar to $K_L \to \mu^+ \mu^-$, BR($K_L \to \pi^0 \, \overline{l} \, l$) is dominated by SM contribution when $\Lambda_{Z'}=\mathcal{O}(10^3)$ TeV.
Then, our predictions are below the experimental bounds.

The LFV decay of $K_L$, $K_L \to \pi^0 \, e^\mp \, \mu^\pm$, is also experimentally constrained as \cite{KTeV:2007}
\beq
{\rm BR} (K_L \to \pi^0 e^\mp \mu^\pm)<7.6 \times 10^{-11}.
\eeq
Using $\Lambda_{Z'}=1.4 \times 10^3$ TeV and typical values of $A^d_{sd}$ and $A^l_{\mu e}$, BR($K_L \to \pi^0 e^- \mu^+$) is
\begin{eqnarray}
{\rm BR}(K_L \to \pi^0 e^- \mu^+) \simeq 2.0 \times 10^{-20} \left(\frac{1400\,{\rm TeV}}{\Lambda_{Z'}}\right)^4 \left(\frac{{\rm Im}(A^d_{sd})}{0.1}\right)^2 \left(\frac{|A^l_{\mu e}|}{0.04}\right)^2.
\end{eqnarray}
Thus, we conclude that our model is not threaten by this process, as far as $\Lambda_{Z'}$ is much bigger than ${\cal O}(10)$ TeV.

\subsubsection{$B_s \to \mu^+ \mu^-$ and $B \to \mu^+ \mu^-$}

In our model, there are large flavor violating $Z'$ couplings in the $(b, \, s)$ and $(b, \, d)$ elements.
Especially, $A^d_{bs}$ tends to be large, as shown in Fig. \ref{fig;FCNC2}. 
Then, the rare $B_s$ decay would constrain our model strongly.

$B_s \to \mu^+ \mu^- $ and $B \to \mu^+ \mu^- $ have been measured at the LHC: BR$(B_s \to \mu^+ \mu^-)=2.8^{+0.7}_{-0.6} \times 10^{-9}$ and BR$(B \to \mu^+ \mu^-)=3.9^{+1.5}_{-1.4} \times 10^{-10}$ \cite{CMS:2014xfa}.
The SM predictions are BR$(B_s \to \mu^+ \mu^-)=(3.66 \pm 0.23) \times 10^{-9}$ and BR$(B \to \mu^+ \mu^-)=(1.06 \pm 0.09) \times 10^{-10}$ \cite{Bobeth:2013uxa}, which are almost consistent with the experimental results,
although the errors are large. 
In our model, the both leptonic decays are deviated from the SM predictions as follows: \cite{Buras:2012jb}
\begin{eqnarray}
\frac{BR(B_s \to \mu^+ \mu^-)}{BR(B_s \to \mu^+ \mu^-)_{\rm SM}}&=& \left |1-  \frac{(C^{l \, B_s}_{L})_{\mu \mu}}{g^2_{SM} \eta_Y Y_0(x_t) (V_{CKM})^*_{tb} (V_{CKM})_{ts} } \right |^2, \label{eq;Bsmumu}
\end{eqnarray}
where $g^2_{SM}=\sqrt{2}G_F \alpha/(\pi \sin^2 \theta_W)$ and $\eta_Y=1.012$ \cite{Buchalla:1998ba} are defined. 
$(C^{l \, B}_{L})_{\mu \mu}$ is given by
replacing $A^d_{sd}$ with $A^d_{bs}$ in $(C^{l }_{L})_{\mu \mu}$. $BR(B \to \mu^+ \mu^-)$ can be also described
by using $A^d_{bd}$ and $(V_{CKM})_{td}$ instead of $A^d_{bs}$ and $(V_{CKM})_{ts}$ in Eq. (\ref{eq;Bsmumu}).
Note that $(C^{l \, B_s}_{L})_{\mu \mu} $ depends on $A^l_{\mu \mu}$ as well.

\begin{figure}[!t]
\begin{center}
{\epsfig{figure=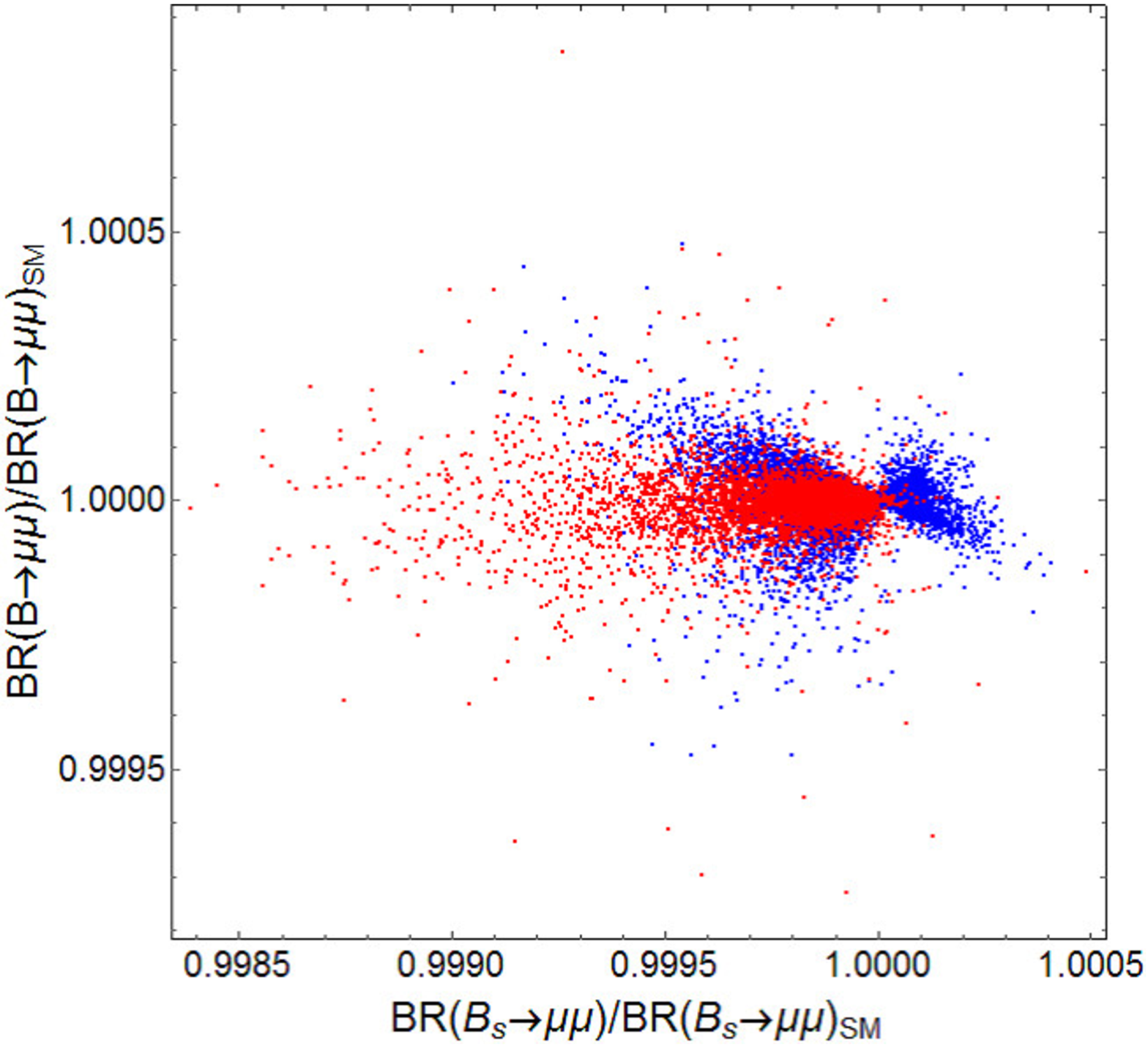,width=0.45\textwidth}}\hspace{0.5cm}{\epsfig{figure=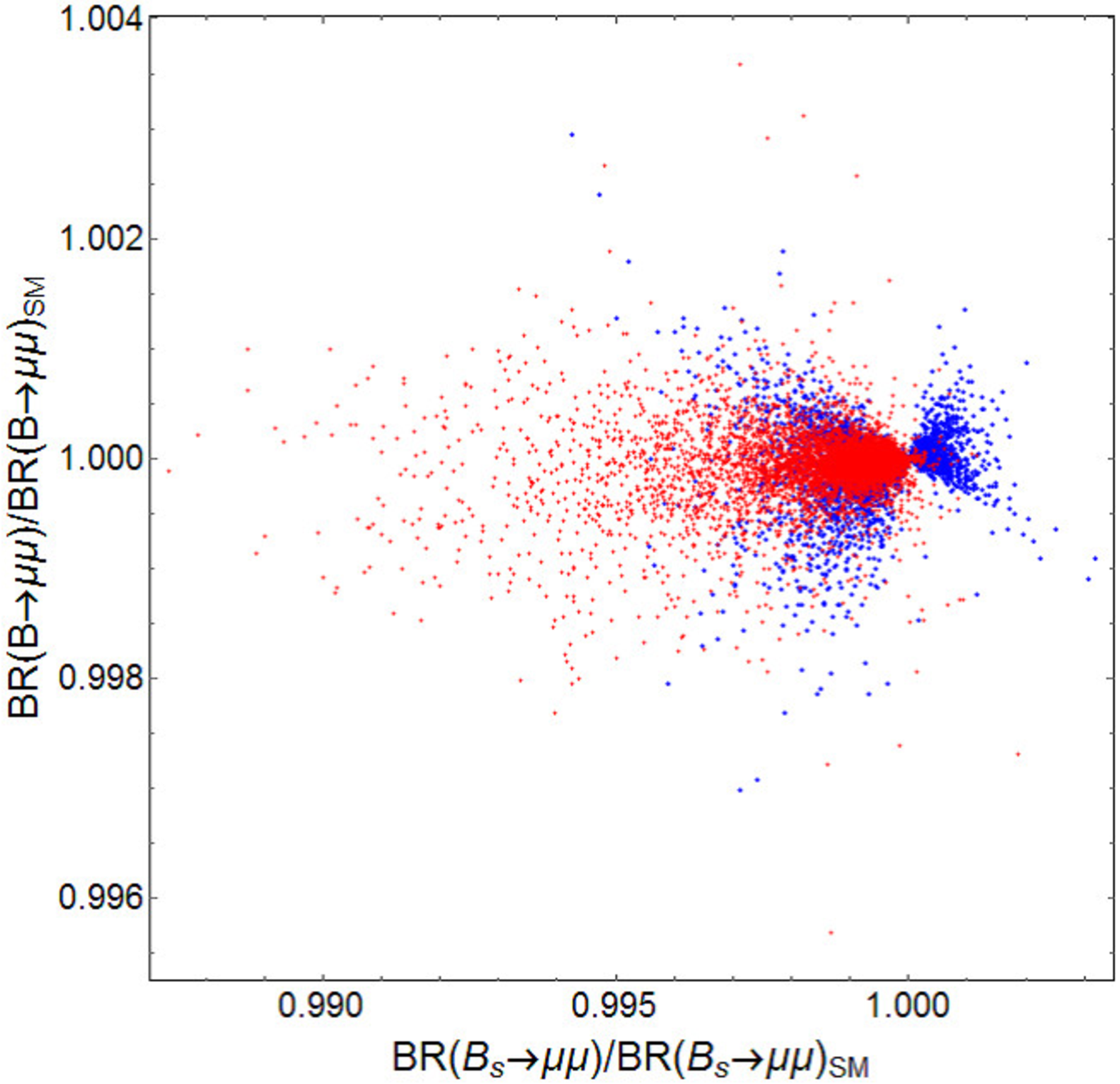,width=0.45\textwidth}}
\caption{ Our predictions for the deviation of BR($B_s \to \mu^+ \mu^-$) and BR($B \to \mu^+ \mu^-$) with $\Lambda_{Z'}=1400$ TeV (left) and $\Lambda_{Z'}=500$ TeV (right). The coefficients of higher-dimensional operators satisfy $|\epsilon \,c^d_{ij}|<10^{-2}$ (red) and $|\epsilon \,c^d_{ij}|<10^{-3}$ (blue). 
In these figures, the constraint, $|\delta \epsilon_K| \leq 0.3$, is assigned.}
\label{fig;Bmumu}
\end{center}
\end{figure}

Fig. \ref{fig;Bmumu} shows our predictions for the deviation of BR($B_s \to \mu^+ \mu^-$) and BR($B \to \mu^+ \mu^-$)
in the each case with $\Lambda_{Z'}=1400$ TeV (left) and $\Lambda_{Z'}=500$ TeV (right).
The deviation of BR($B_s \to \mu^+ \mu^-$) is large, compared to the one of BR($B \to \mu^+ \mu^-$),
but it is at most a few \% even in the $500$ TeV $\Lambda_{Z'}$ case.

\subsection{Flavor violating processes in $\mu$ decay  }
\label{sec3-2}
The tree-level FCNCs of $Z'$ predict the LFV decays.
Depending on the sizes of the coefficients of higher-dimensional operators, all elements of the FCNCs could be ${\cal O}(1)$, and then
the LFV processes, which face the stringent experimental constraints, are
 important in our model; that is, $\mu \to 3 \, e$ and $\mu$-$e$ conversions in nuclei
should be taken into account. Note that $\mu \to e \gamma$ is one of the relevant processes, but
it is suppressed in our model, because of the heavy $Z'$ and the loop suppression.

\subsubsection{$\mu \to 3 \, e$}
 
First, let us discuss $\mu \to 3 \, e$ in our model. The LFV is caused by the following 4-Fermi interactions:
\beq
{\cal H}^{ \mu \to 3e}= C^{3e}_{L} (\overline{e_L} \gamma^\mu \mu_L)(\overline{e_L} \gamma^\mu e_L)+C^{3e}_{R} (\overline{e_L} \gamma^\mu \mu_L)(\overline{e_R} \gamma^\mu e_R), \label{eq;mu3e4fermi}
\eeq
where the coefficients are given by
\begin{eqnarray}
 C^{3e}_{L}  &=& A^l_{e \mu} \left \{ \frac{A^l_{ee}}{\Lambda^2_{Z'}}  - \frac{ \cos 2 \theta_W }{2} \frac{1}{\Lambda^2_Z} \right \}, \\
  C^{3e}_{R} &=& A^l_{e \mu} \left \{  \frac{1}{\Lambda^2_{Z'}}  +  \sin^2 \theta_W  \frac{1}{\Lambda^2_{Z}}  \right \}.
\end{eqnarray}
The branching ratio of $\mu \to 3 \, e$ can be evaluated, ignoring the $Z'$ contribution to $\mu \to e \overline{\nu} \nu$:

\begin{eqnarray}
\text{BR}(\mu \rightarrow 3 \, e)&=&\frac{m_{\mu}^5}{1536 \, \pi^3 \,  \Gamma_{\mu}  }\left(2 \left| C^{3e}_{L} \right|^2 +\left| C^{3e}_{R} \right|^2\right)  \label{eq;mu3eBR} \\
&\simeq& 5.8\times 10^{-18} \left(\frac{1400\,{\rm TeV}}{\Lambda_{Z'}}\right)^4 \left(\frac{|A^l_{\mu e}|}{0.04}\right)^2, \label{eq;mu3eBRtypical}
\end{eqnarray}
where $m_{\mu}$ and $\Gamma_{\mu}$ are mass and total decay width for $\mu$, respectively. 

\begin{figure}[!t]
\begin{center}
{\epsfig{figure=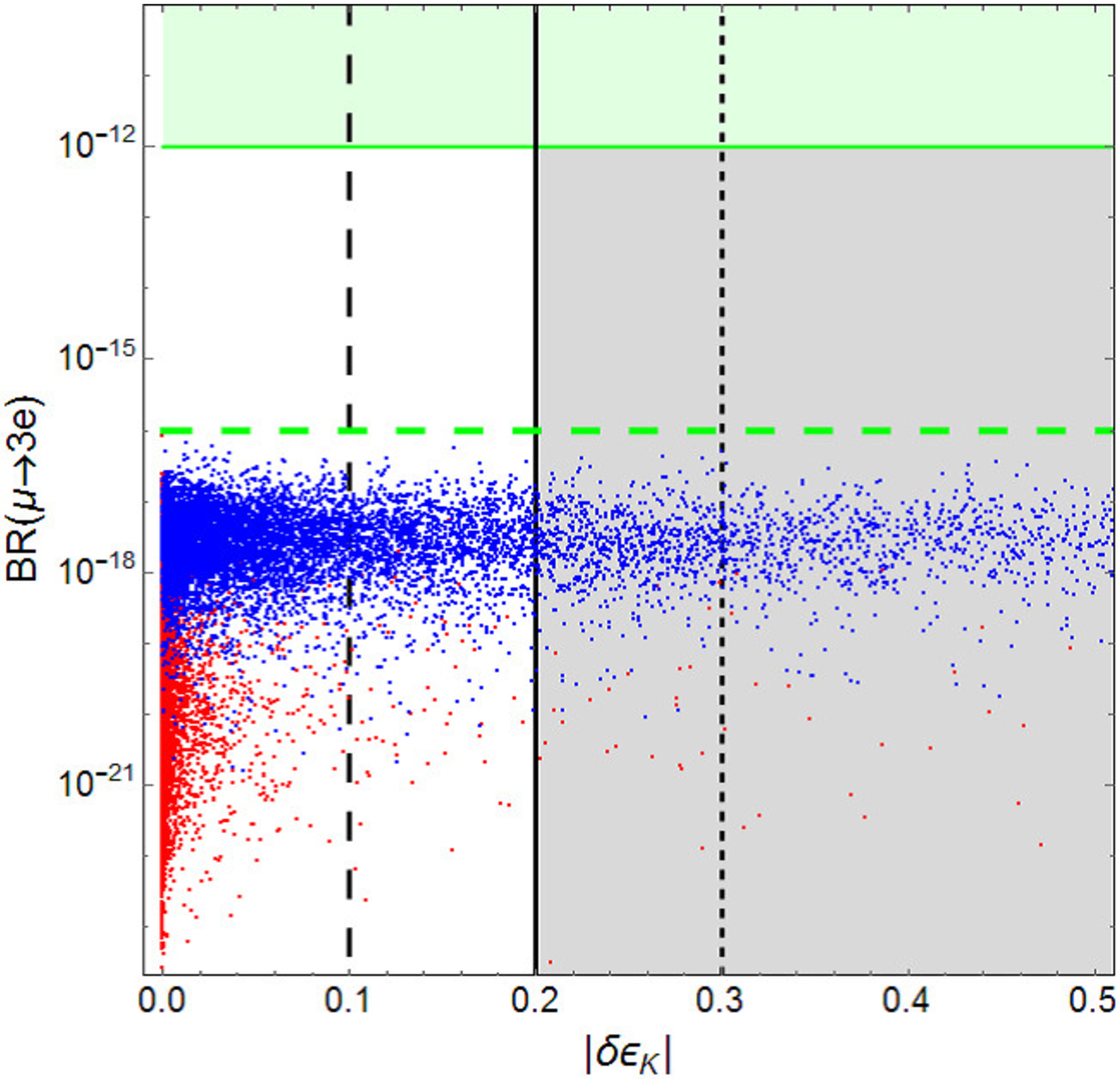,width=0.45\textwidth}}\hspace{0.5cm}{\epsfig{figure=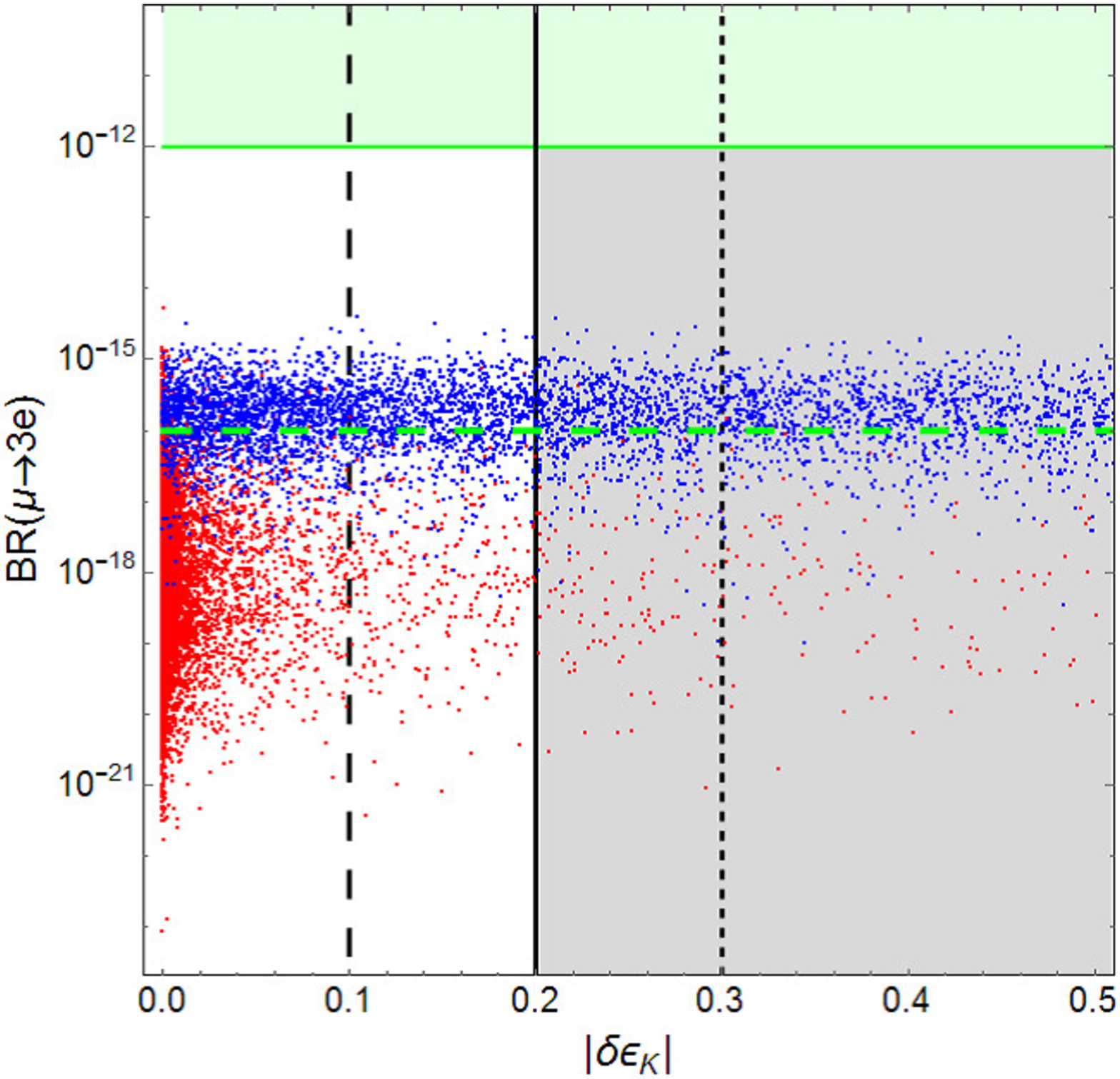,width=0.45\textwidth}}
\caption{ Our predictions for the deviation of BR($\mu \rightarrow 3 \, e$) with $\Lambda_{Z'}=1400$ TeV (left) and $\Lambda_{Z'}=500$ TeV (right). The coefficients of higher-dimensional operators satisfy $|\epsilon \,c^d_{ij}|<10^{-2}$ (red) and $|\epsilon \,c^d_{ij}|<10^{-3}$ (blue). The green region is excluded by the SINDRUM experiment \cite{Bellgardt:1987du} and the green dashed line is the future prospected bound \cite{Blondel:2013ia}.}
\label{fig;mu3e}
\end{center}
\end{figure}

This LFV process has been investigated at the SINDRUM experiment: BR$(\mu \rightarrow 3 \,e) <1.0 \times 10^{-12}$ \cite{Bellgardt:1987du}. The coming experiment will reach $\mathcal{O}(10^{-16})$ \cite{Blondel:2013ia}.
Fig. \ref{fig;mu3e} shows the correlation between $\delta (\epsilon_K)$ and BR$(\mu \to 3 \, e)$, setting $\Lambda_{Z'}=1400$ TeV (left) and $\Lambda_{Z'}=500$ TeV (right).
The green region is excluded by the SINDRUM experiment \cite{Bellgardt:1987du} and the dashed green line corresponds to the expected upper bound in the Mu3e experiment \cite{Blondel:2013ia}.
According to the figures, we can expect that BR($\mu \to 3e$) is less than ${\cal O}(10^{-15})$,
as far as $\Lambda_{Z'}>500$ TeV. 
When $\Lambda_{Z'}$ is $500$ TeV which correspond to $M_{Z'} \simeq 36$ TeV, BR($\mu \to 3e$) is about $3.5 \times 10^{-16}$ and can exceed the future sensitivity. Note that $|\delta \epsilon_K|$ is also enhanced in this case, as shown in Fig. \ref{fig;KKbarmixing}.

 \subsubsection{$\mu$-$e$ conversion}
 The $\mu$-$e$ conversions in nuclei are also predicted by our $Z'$ interaction.
 Now, we assume that the coherent conversion, in which the final state is the same as the initial, is dominant and then we concentrate on the contributions derived from the operators,
 \beq
{\cal H}^{\mu \mathchar`-e} =C_{q}^{\mu \mathchar`-e  }(\overline{q} \gamma_\mu q)(\overline{e_L} \gamma^\mu \mu_L),
\eeq
where the coefficients are given by
\begin{eqnarray}
C_u^{\mu \mathchar`-e }&=&A_{e \mu }^{l} \left \{ \left ( \frac{1}{4} -\frac{2}{3} \sin^2 \theta_W \right )  
\frac{1}{\Lambda^2_Z} \right \},  \\
 C_d^{\mu \mathchar`-e}&=&-A_{e \mu}^{l}\left \{\frac{A^d_{dd}+1}{2\Lambda^2_{Z'}}    +   \left ( \frac{1}{4} -\frac{1}{3} \sin^2 \theta_W \right ) \frac{1}{\Lambda^2_Z}  \right \}. 
\end{eqnarray}
 The conversion rate of muon $\omega_{\text{conv}}$ is 
\begin{equation}
\omega_{\text{conv}}=4 m_\mu^5
\left| \left( 2 C_u^{\mu \mathchar`-e} + 
C_d^{\mu \mathchar`-e} \right)V^{(p)}+
\left( C_u^{ \mu \mathchar`-e} + 
2 C_d^{\mu \mathchar`-e} \right)V^{(n)}\right|^2,
\end{equation}
where $V^{(p)}$ and $V^{(n)}$ are overlap integrals which 
depend on the nucleus species.
The branching ratio of the $\mu$-$e$ conversion is
\begin{eqnarray}
\text{BR}(\mu \, N\rightarrow e \, N) &=& \frac{\omega_{\text{conv}}}{\omega_{\text{capt}}} \nonumber \\
&\simeq& 4.0 \times 10^{-17}\, (1.4 \times 10^{-17}) \left(\frac{1400\,{\rm TeV}}{\Lambda_{Z'}}\right)^4 \left(\frac{|A^l_{\mu e}|}{0.04}\right)^2,
\end{eqnarray}
where $\omega_{\text{capt}}$ is the muon capture rate.
The overlap integrals $V^{(p)}$ and $V^{(n)}$ and 
the muon capture rate $\omega_{\text{capt}}$ have been calculated 
in Ref. \cite{Kitano:2002mt} for the each nucleus species.
We also show the typical value of BR($\mu  \, {\rm Au} \to e  \, {\rm Au}$) (BR($\mu  \, {\rm Al} \to e  \, {\rm Al}$)) in our model.

\begin{figure}[!t]
\begin{center}
  {\epsfig{figure=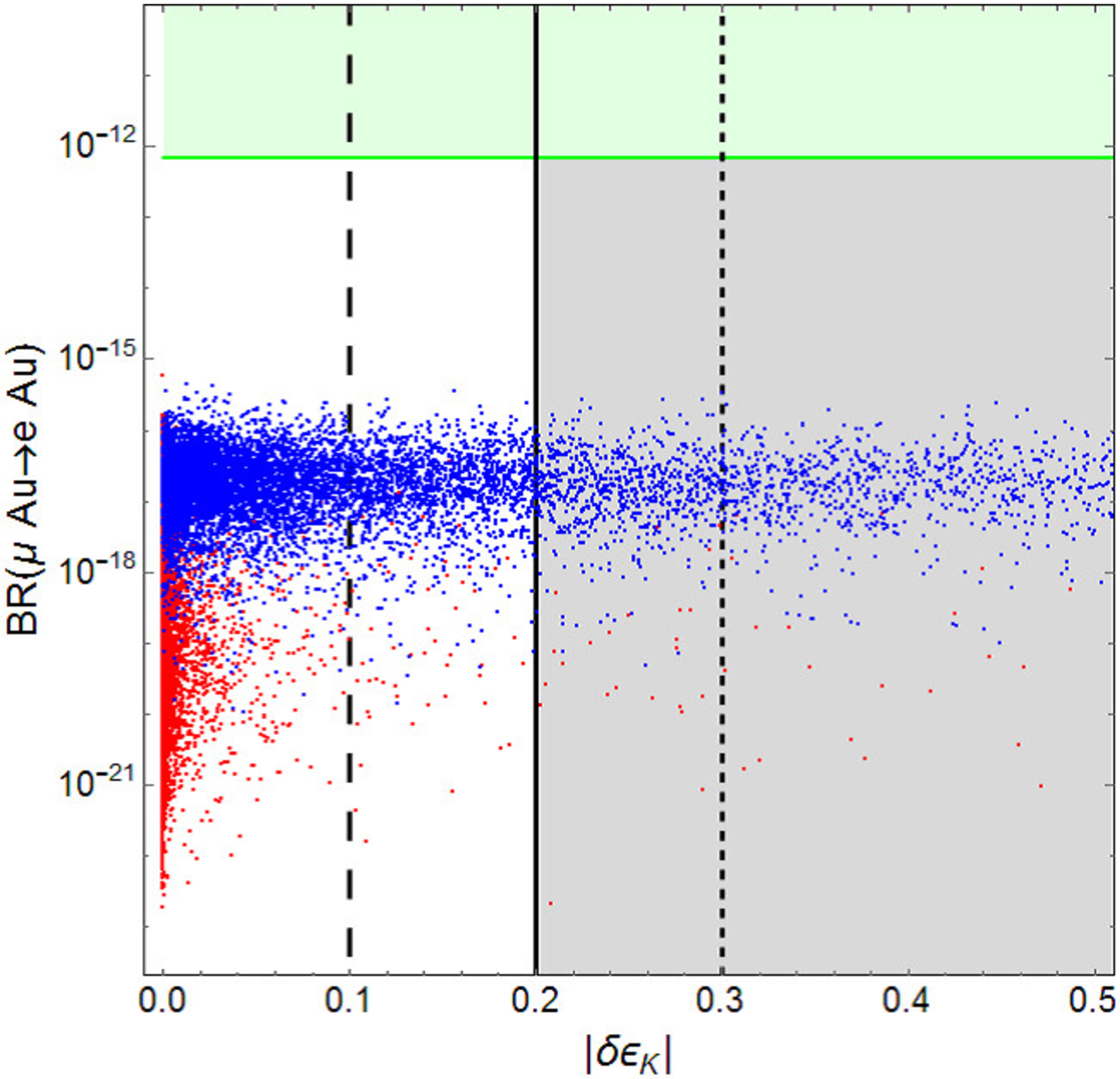,width=0.45\textwidth}}\hspace{0.5cm}{\epsfig{figure=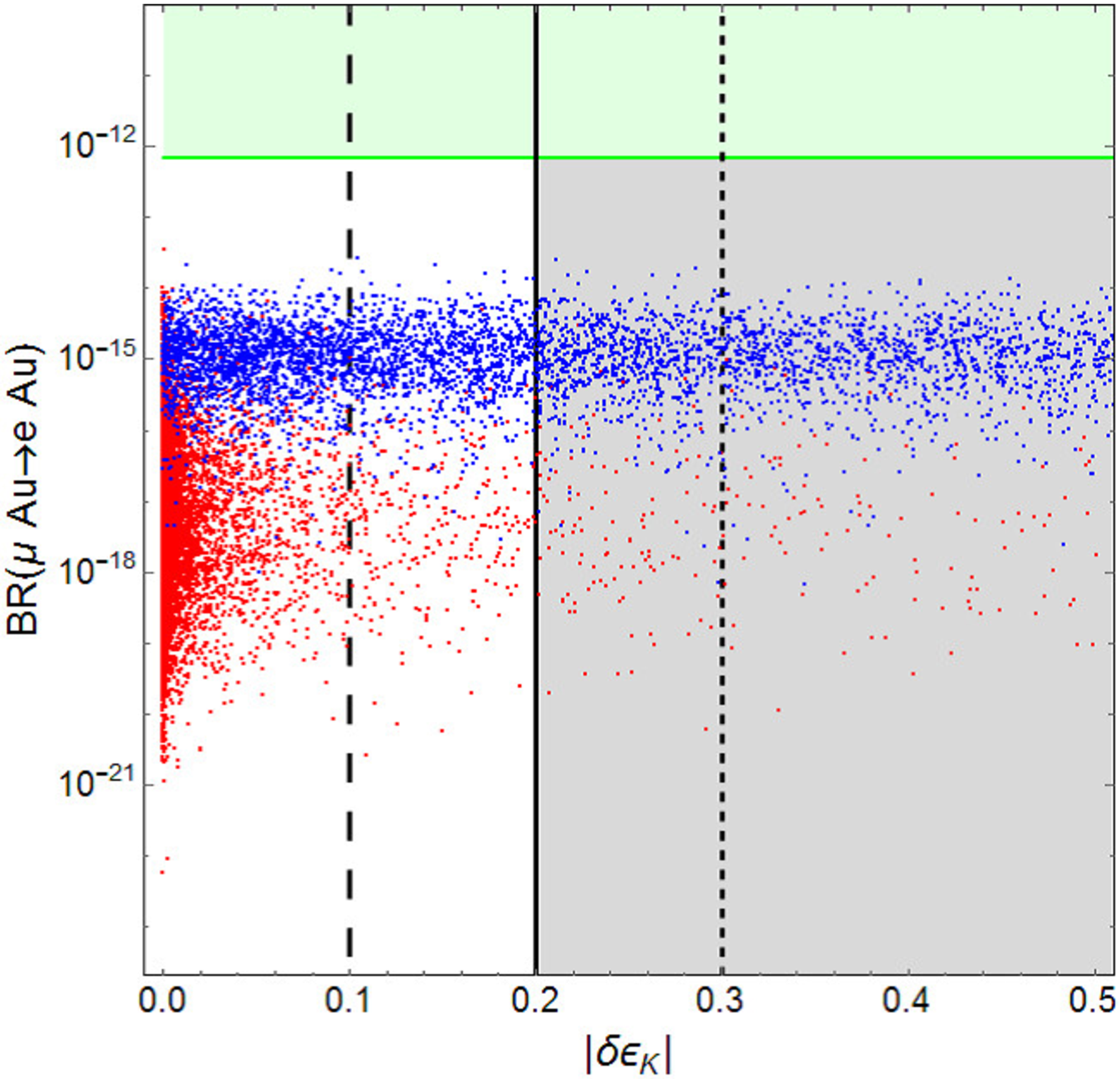,width=0.45\textwidth}}
  \vspace{0.5cm}
  {\epsfig{figure=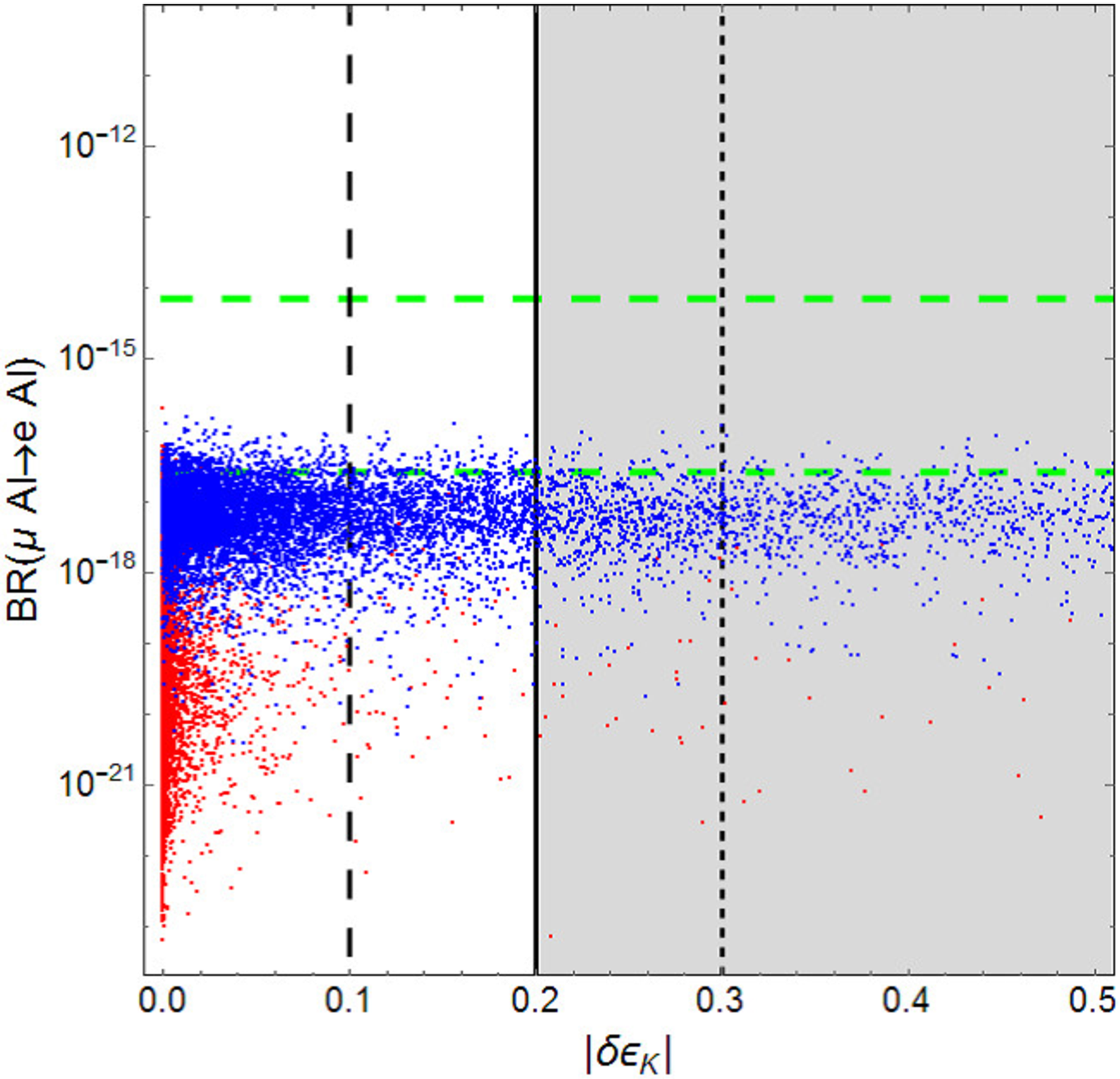,width=0.45\textwidth}}\hspace{0.5cm}{\epsfig{figure=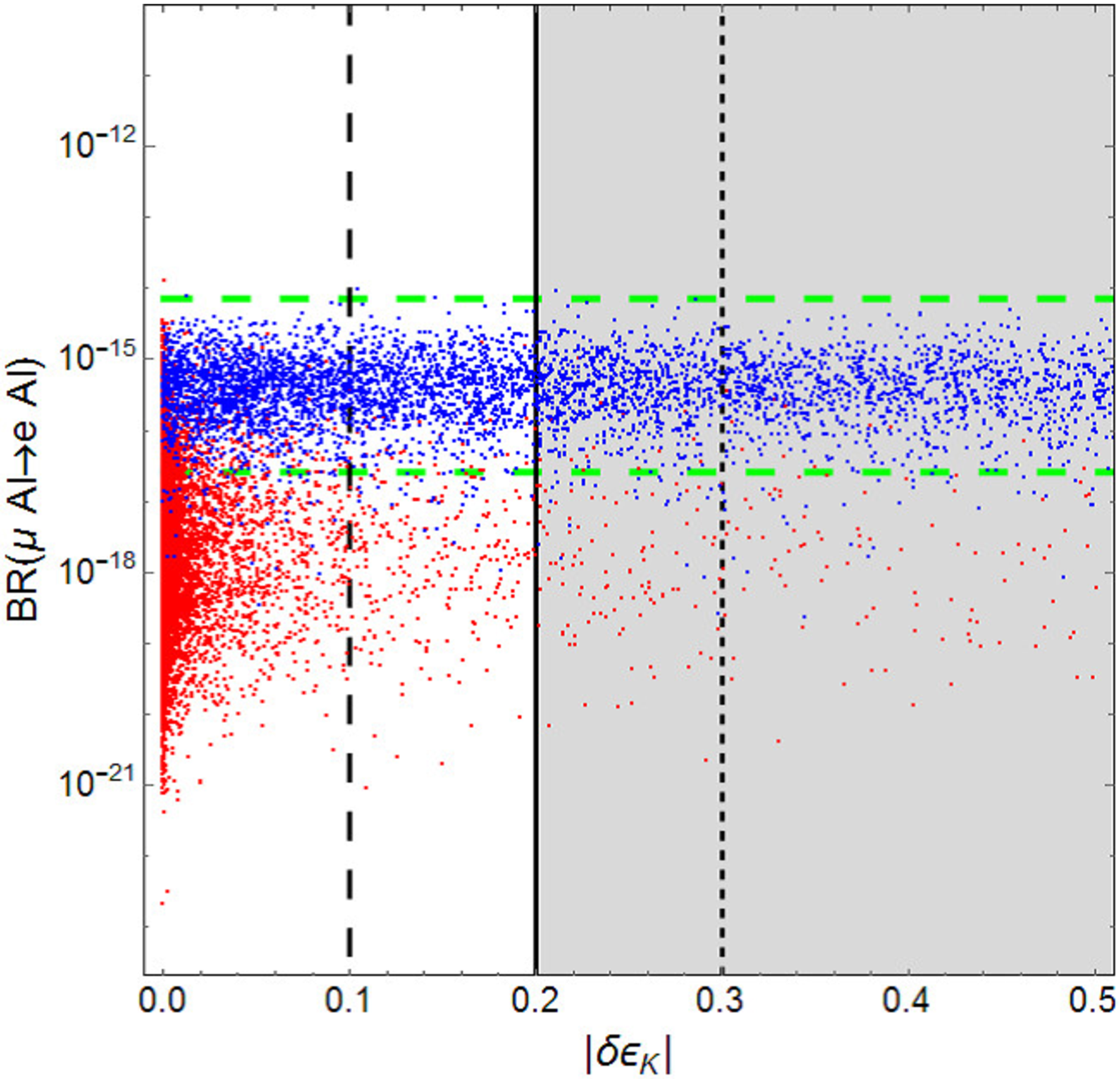,width=0.45\textwidth}}
\caption{ Our predictions for BR($\mu  \, {\rm Au} \to e  \, {\rm Au}$) (upper panels) and BR($\mu \, {\rm Al} \to e \,  {\rm Al}$) (lower panels). We set $\Lambda_{Z'}=1400$ TeV in left two panels and $\Lambda_{Z'}=500$ TeV in right two panels. The coefficients of higher-dimensional operators satisfy $|\epsilon \,c^d_{ij}|<10^{-2}$ (red) and $|\epsilon \,c^d_{ij}|<10^{-3}$ (blue). In the upper panels, green region shows the experimental bound \cite{Bertl:2006up}. In the lower pannels, two green dashed lines show future sensitivity from COMET-I (upper one) and COMET-II (lower one) experiment \cite{Kuno:2013mha,COMET:2014}.}
\label{fig;mueconversion}
\end{center}
\end{figure}
 %%%%%%%%%%%%%%%%%%%%%%%%%%%%%%%%%%%%%%%
%%%%%%%%%%%%%%%%%%%%%%%%%%%%%%%%%%%%%%%
Fig. \ref{fig;mueconversion} shows the correlations on $\delta \epsilon_K$ and the $\mu$-$e$ conversions.
The green region is excluded by the SINDRUM experiment \cite{Bertl:2006up}. The dashed green lines are the future prospects for BR($\mu \, {\rm Al} \to e \, {\rm Al}$).
In these observables, the upper limits are depicted in Fig. \ref{fig;mueconversion}, depending on the sizes of $\epsilon$
and $\Lambda_{Z'}$: BR($\mu \, N \to e \, N) < {\cal O}(10^{-15})$.
Although these results are much below the current experimental limit, there is a chance to reach the future sensitivity of the COMET-II experiment \cite{Kuno:2013mha,COMET:2014} in the mode of $\mu \, {\rm Al} \to e \, {\rm Al}$: BR($\mu \, {\rm Al} \to e \, {\rm Al}) \simeq 10^{-15}$ when $\Lambda_{Z'}$ is set to $500$ TeV.

\subsection{Contributions to LFV $\tau$ decays}
\label{sec3-4}
Finally, let us discuss LFV $\tau$ decays: $\tau \to l_i \, l_j \, \bar{l}_k$ and $\tau \to l_i \, P^0$, where $P^0$ denotes neutral mesons, $P^0=\pi^0, \, K_S$, in this section. 
To begin with, we discuss the leptonic decay, $\tau \to l_i \, l_j \, \bar{l}_k$. This decay is caused by the following 4-Fermi interactions similar to Eq. (\ref{eq;mu3e4fermi}):
\beq
{\cal H}^{ \tau \to 3l}= {C^{3l}_{L}}_{ijk} (\overline{l_L}_i \gamma^\mu \tau_L)(\overline{l_L}_j \gamma_\mu {l_L}_k)+{C^{3l}_{R}}_{ijk} (\overline{l_L}_i \gamma^\mu \tau_L)(\overline{l_R}_j \gamma^\mu {l_R}_k),
\eeq
where the coefficients are given by
\begin{eqnarray}
 {C^{3l}_{L}}_{ijk} &=& A^l_{i \tau} \left \{ \frac{A^l_{jk}}{\Lambda^2_{Z'}}  - \frac{ \cos 2 \theta_W }{2} \frac{\delta_{jk}}{\Lambda^2_Z} \right \}, \\
 {C^{3l}_{R}}_{ijk} &=& A^l_{i \tau} \delta_{jk} \left \{  \frac{1}{\Lambda^2_{Z'}}  +  \sin^2 \theta_W  \frac{1}{\Lambda^2_{Z}}  \right \}.
\end{eqnarray}
In LFV $\tau$ decays, there are many modes, e.g. $\tau \to 3 \mu$, $\tau \to\mu^- e^+ e^- $, $\tau \to e^+ \mu^- \mu^-$ and so on. The branching ratios for some of these modes can be estimated by changing $m_{\mu} \to m_{\tau}$, $\Gamma_{\mu} \to \Gamma_{\tau}$ and $C_{L,R}^{3e} \to {C_{L,R}^{3l}}_{ijk}$ in Eq. (\ref{eq;mu3eBR}). In the case that there are three different charged leptons in final state, the branching ratio for $\tau \to l_i l_j \bar{l}_k$ is \cite{Langacker:2000}
\begin{eqnarray}
{\rm BR}(\tau \to l_i \, l_j \,  \bar{l}_k) = \frac{m_{\tau}^5}{1536  \, \pi^3  \, \Gamma_{\tau} }\left(\left| {C_{L}^{3l}}_{ijk} + {C_{L}^{3l}}_{jik} \right|^2 + \left| {C_{R}^{3l}}_{ijk} \right|^2 + \left| {C_{R}^{3l}}_{jik} \right|^2 \right).
\end{eqnarray}
We show the typical values of branching ratios for all decay modes of  $\tau \to l_i \, l_j \, \bar{l}_k$ in Table \ref{tab;taudecay} and we found that these modes are extremely smaller than the experimental bounds \cite{PDG,Hayasaka:2010np} in our model.

Next, let us discuss $\tau \to l_i \, \pi^0$ and  $ \tau \to l_i \, K_S$. These decays are caused by the following interactions:
\beq
{\cal H}^{ \tau \to l P^0}= C^{lP^0}_{L\,ijk} (\overline{l_L}_i \gamma^\mu \tau_L)(\overline{q_L}_j \gamma_\mu {q_L}_k)+C^{lP^0}_{R\,ijk} (\overline{l_L}_i \gamma^\mu \tau_L)(\overline{q_R}_j \gamma_\mu {q_R}_k),
\eeq
where the coefficients are similar to Eq. (\ref{eq;deltaS1-1}):
\begin{eqnarray}
C^{lP^0}_{I\,ijk} = A^l_{i\tau}  \left \{ \frac{ (Q^q_{I})_{jk} }{ \Lambda^2_{Z'} }+  \frac{ \delta_{jk} }{\Lambda^2_{Z} } \left ( \tau^q_I - Q_e^{q} \sin^2 \theta_W \right )  \right \}.
\end{eqnarray}
The branching ratios of $\tau \to l_i \,  \pi^0$ and $\tau \to l_i \, K_s^0$ are evaluated by the following expression \cite{Langacker:2000}:
\begin{eqnarray}
&&{\rm BR}(\tau \to l_i \, \pi^0) =  \frac{{\rm BR}(\tau \to \nu_{\tau} \pi^-)}{16 \, |(V_{CKM})_{ud}|^2 G_F^2} \times \left(|C^{lP^0}_{L\,iuu} - C^{lP^0}_{R\,iuu} - C^{lP^0}_{L\,idd} + C^{lP^0}_{R\,idd}|^2\right), \\
&&{\rm BR}(\tau \to l_i \, K_S) =  \frac{{\rm BR}(\tau \to \nu_{\tau} K^-)}{16  \,  |(V_{CKM})_{us}|^2 G_F^2} \times \left(|C^{lP^0}_{R\,isd} - C^{lP^0}_{R\,ids}|^2\right),
\end{eqnarray}
where BR($\tau \to \nu_{\tau} \pi^-$)=0.1083 and BR($\tau \to \nu_{\tau} K^-$)=0.007 \cite{PDG}.

We summarize the typical values of each branching ratio for $\tau \to l_i \, P^0$ and each experimental bound \cite{PDG} in Table \ref{tab;taudecay}. We see that these decay modes are also smaller than the experimental bounds.
Note that BR($\tau \to e \, P^0$) is smaller than BR($\tau \to \mu \, P^0$) because this type of branching ratio is proportional to $|A^l_{i\tau}|^2$ and roughly speaking, $|A^l_{e\tau}|<|A^l_{\mu\tau}|$.

\begin{table}[!t]
\begin{center}
  \begin{tabular}{|c|c|c|} \hline
    $\tau$ decay mode & value of BR & exp. bound ($\times 10^{-8}$) \cite{PDG,Hayasaka:2010np} \\ \hline
    $e^- e^+ e^-$ & $1.2 \times 10^{-18}$ & $< 2.7$\hspace{1.0em} \\
    $e^- \mu^+ \mu^-$ & $4.2 \times 10^{-19}$ & $< 2.7$\hspace{1.0em} \\
    $e^+ \mu^- \mu^-$ & $1.5 \times 10^{-18}$ & $< 1.7$\hspace{1.0em} \\
    $\mu^- e^+ e^-$ & $3.7 \times 10^{-15}$ & $< 1.8$\hspace{1.0em} \\
    $\mu^+ e^- e^-$ & $2.8 \times 10^{-22}$ & $< 1.5$\hspace{1.0em} \\
    $\mu^- \mu^+ \mu^-$ & $2.7 \times 10^{-15}$ & $< 2.1$\hspace{1.0em} \\ \hline
    $e^- \pi^0$ & $2.2 \times 10^{-19}$ & $< 8.0$\hspace{1.0em} \\
    $\mu^- \pi^0$ & $1.2 \times 10^{-15}$ & $< 11$\hspace{1.0em} \\
    $e^- K_s^0$ & $1.2 \times 10^{-21}$ & $< 2.6$\hspace{1.0em} \\
    $\mu^- K_s^0$ & $6.6 \times 10^{-18}$ & $< 2.3$\hspace{1.0em} \\ \hline
  \end{tabular}
  \caption{ The typical values of each $\tau$ decay mode. In this table, we use $\Lambda_{Z'}=1.4 \times 10^3$ TeV and typical values of $A^d_{ij}$ and $A^l_{ij}$.}
  \label{tab;taudecay}
\end{center}
\end{table}
%}

\section{Summary}
\label{sec5}
The grand unification is one of the attractive hypotheses to solve the mystery of our nature.
The SO(10) GUT elegantly explains the origin of the SM gauge groups and the minimal setup
shows that all matters except Higgs fields can be unified into a ${\bf 16}$-representational field
in the each generation surprisingly. 
Our nature, however, is not so simple. 
The hierarchical structure of the fermions exists in the each sector, (i.e. up-type, down-type, and leptonic Yukawa couplings), but the observed values unfortunately seem to dislike the unification of the Yukawa couplings.
In Ref. \cite{Hisano-so10}, we propose a SO(10)-GUT model, introducing ${\bf 10}$-representational matter fields,
in order to realize the realistic Yukawa couplings. In this model, the SM fields are given by the linear combination
of the parts of the ${\bf 10}$- and ${\bf 16}$-representational fields, and especially the mass hierarchy between top and bottom quarks is achieved by the mixing. Although we have to expect additional contributions such as higher-dimensional operators to the fermion mass matrices, we have successfully reproduced the realistic Yukawa couplings in this paper. 

The important and interesting feature of our model is to predict the flavor violating couplings of $Z'$.
SO(10) predicts an extra U(1)$^\prime$ symmetry. In our scenario, the matter fields are given by the two different fields of SO(10), which carry different U(1)$^\prime$ charges. Then, the flavor violating $Z'$ interaction is induced by the spontaneous U(1)$^\prime$ symmetry breaking, and we can expect that the $Z'$ couplings are related to the Yukawa couplings, such as the mass hierarchy and the mixing. In fact, we find that the flavor violating $Z'$ couplings, denoted by $A^{d,l}_{ij}$, depend on the fermion masses and the CKM matrix, and we derive the explicit forms of $A^{d,l}_{ij}$, although the unknown parameters appear according to the higher-dimensional operators. Interestingly, we see that there are some correlations among the flavor violating $Z'$ couplings.
For example, $A^{d}_{ij}$ ($A^{l}_{ij}$) are linear to $m^d_i$ and $m^d_j$ ($m^l_i$ and $m^l_j$), so that
$A^{d}_{bs}$ tends to be large and $A^{l}_{ij}/A^{d}_{ij}$ is approximately estimated as $m^l_im^l_j/(m^d_im^d_j)$,
in the limit that $\epsilon \to 0.$

In this paper, we especially investigate the flavor physics relevant to our FCNCs.
$A^{d,l}_{ij}$, actually, could be ${\cal O}(1)$, depending on the size of the coefficients of higher-dimensional operators.
Then, $\epsilon_K$ is the most sensitive to our model.
Besides, the large $(b,\,s)$ element of the $Z^\prime$ coupling predicts relatively large deviations of $\Delta M_{B_s}$ and 
$B_s$ decay.

Moreover, we find that there are correlations between the flavor violation in the quark sector and LFV.
In the LFV, the stringent constraints come from the LFV $\mu$ decays, such as $\mu$-$e$ conversion and $\mu \to 3e$.
They are expected to be developed near future, so that our model could be tested, for instance, in the COMET and Mu2e experiments.
As we see Fig. \ref{fig;mueconversion}, our prediction could reach the future prospect of the COMET without conflict with $\epsilon_K$, if $Z'$ scale is ${\cal O}(100)$ TeV. 
Other future experiments for $\mu$-$e$ conversion are planned 
\cite{Natori:2014yba,Mu2eproposal,PRISM}, and
our model can be tested if their sensitivities reach $O(10^{-15})$.
If we assume that the extra U(1)$^\prime$ is radiately broken around the SUSY scale, $Z'$ scale would be ${\cal O}(100)$ TeV to realize 125 GeV Higgs in the high-scale SUSY scenario.
Then, it is implied that our SUSY model can be tested indirectly, even though the
SUSY scale is much higher than the energy scale reached by the LHC.

Before closing our discussion, let us give some comments on the other observables in flavor physics.
In our model, all elements of the tree-level FCNCs involving $Z'$ could be large in principle, so that
all observables may be relevant to our model. 
One of the processes that recently attract attention is the direct CP violation in $K \to \pi \pi$.
As pointed out in Ref. \cite{Bai:2015nea,Buras:2015yba}, the SM prediction of $\epsilon^\prime/\epsilon$
is deviated from the experimental results, according to the lattice QCD calculation.
Another interesting process would be $b \to s$ transition, such as $B \to K \,l \,l$, which is slightly deviated from the SM prediction \cite{Aaij:2013qta,Aaij:2014ora}.
The new physics interpretations are given by, e.g. Ref. \cite{Descotes-Genon:2013wba}, and summarized in Refs. \cite{Altmannshofer:2014rta,Descotes-Genon:2015uva}. In those processes, our predictions will depart from the SM predictions as well, so that it would be interesting to discuss if our model can resolve the discrepancies,
although relatively low $Z'$ scale should be assumed. This work may be done elsewhere in the future.

\section*{Acknowledgments}
This work is supported by Grant-in-Aid for Scientific research from
the Ministry of Education, Science, Sports, and Culture (MEXT), Japan,
No.16H00867 and 16H06492 (for J.H.), and National Research Foundation of
Korea (NRF) Research Grant NRF- 2015R1A2A1A05001869 (for Y.M.). The
work of J.H. is also supported by World Premier International Research
Center Initiative (WPI Initiative), MEXT, Japan.
The work of Y.S. is supported by the Japan Society for the Promotion of Science (JSPS) Research Fellowships for Young Scientists, No. 16J08299.

\appendix

\section{RG equations for the $\Delta F=2$ processes}
\label{sec;appendix1-2}
The one-loop RG equation for $\widetilde Q^q_1$ is given by
\begin{equation}
\mu \frac{d}{d \mu} \widetilde{C}^q_1 = -\frac{\alpha_s}{2 \pi} \left (\frac{3}{N_c}-3  \right) \widetilde{C}^q_1.
\end{equation}
Using the one-loop description of the RG running of $\alpha_s$, we can estimate the one-loop Wilson coefficients in the each process: for the $K$-$\overline{K}$ mixing,
\beq
\widetilde{C}^K_1(m_K)= \left( \frac{\alpha_s(m_c)}{\alpha_s(m_K)} \right )^{\frac{2}{9}} \left( \frac{\alpha_s(m_b)}{\alpha_s(m_c)} \right )^{\frac{6}{25}}  \left( \frac{\alpha_s(m_t)}{\alpha_s(m_b)} \right )^{\frac{6}{23}} \left( \frac{\alpha_s(M_{Z'})}{\alpha_s(m_t)} \right )^{\frac{2}{7}} \widetilde{C}^K_1(M_{Z'}),
\eeq
and for the $B_{(s)}$-$\overline{B_{(s)}}$ mixing,
\beq
\widetilde{C}^{B_{(s)}}_1(m_b)=   \left( \frac{\alpha_s(m_t)}{\alpha_s(m_b)} \right )^{\frac{6}{23}} \left( \frac{\alpha_s(M_{Z'})}{\alpha_s(m_t)} \right )^{\frac{2}{7}} \widetilde{C}^{B_{(s)}}_1(M_{Z'}).
\eeq

\section{Functions}
\label{sec;appendix1}
The functions which appear in the $K$-$\overline{K}$ and  $B_{(s)}$-$\overline{B_{(s)}}$ mixing are given by
\begin{eqnarray}
S_0(x)&=&  \frac{4x -11x^2+x^3}{4(1-x)^2} - \frac{3x^3 \log x}{2(1-x)^3}, \\
S(x,y)&=&\frac{-3xy}{4(y-1)(x-1)} - \frac{xy(4-8y+y^2) \log y}{4(y-1)^2(x-y)}  \nonumber \\
&&+\frac{xy(4-8x+x^2) \log x}{4(x-1)^2(x-y)}.
\end{eqnarray}
The function for the short-distance contribution to $K_L \to \pi \ov{\nu} \nu$ is defined as
\begin{eqnarray}
X(x)&=& \frac{x}{8} \left \{ \frac{x+2}{x-1} + \frac{3x-6}{(x-1)^2} \log x \right \}.
\end{eqnarray}
The function for $B_{s(d)} \to \mu^+ \mu^-$ is defined as
\begin{eqnarray}
Y_0(x)&=& \frac{x}{8} \left \{ \frac{x - 4}{x - 1} + \frac{3 x}{(x - 1)^2} \ln x \right \}.
\end{eqnarray}

\end{document}